\documentclass{article}
\usepackage{arxiv}
\usepackage{amsmath,amssymb,amsfonts,amsthm}
\usepackage{siunitx}
\usepackage{booktabs}
\usepackage{multirow}
\usepackage{graphicx}
\usepackage{xcolor}
\usepackage{microtype}
\usepackage{multicol}
\usepackage[ruled]{algorithm2e}
\usepackage{hyperref}

\theoremstyle{remark}

\renewcommand{\vec}[1]{\boldsymbol{#1}}

\newcommand{\dd}{\mathrm{d}}
\newcommand{\pfrac}[2]{\frac{\partial #1}{\partial #2}}

\graphicspath{{img}}

\title{A highly efficient computational approach for part-scale microstructure predictions in Ti-6Al-4V additive manufacturing}

\newcommand{\shorttitle}{Highly efficient computational approach for part-scale microstructure predictions in Ti-6Al-4V additive manufacturing}

\author{Sebastian D.~Proell\thanks{corresponding author} \\
	Institute for Computational Mechanics\\
	Technical University of Munich\\
	85748 Garching b. München\\
	\texttt{sebastian.proell@tum.de} \\
	\And
    Julian Brotz\\
    Institute for Computational Mechanics\\
	Technical University of Munich\\
	85748 Garching b. München\\
	\AND
	Martin Kronbichler\\
	Faculty of Mathematics\\
	Ruhr University Bochum\\
	44780 Bochum\\
	\AND
	Wolfgang A.~Wall\\
	Institute for Computational Mechanics\\
	Technical University of Munich\\
	85748 Garching b. München\\
	\And
	Christoph Meier\\
	Institute for Computational Mechanics\\
	Technical University of Munich\\
	85748 Garching b. München
}
\date{\today}

\begin{document}

\maketitle

\begin{abstract}
Fast and efficient simulations of metal additive manufacturing (AM) processes are highly relevant to exploring the full potential of this promising manufacturing technique. The microstructure composition plays an important role in characterizing the part quality and deriving mechanical properties. When complete parts are simulated, one often needs to resort to strong simplifications such as layer-wise heating due to the large number of simulated time steps compared to the small time step sizes. This article proposes a scan-resolved approach to the coupled thermo-microstructural problem. Building on a highly efficient thermal model, we discuss the implementation of a phenomenological microstructure model for the evolution of the three main constituents of Ti-6Al-4V: stable $\alpha_s$-phase, martensite $\alpha_m$-phase and $\beta$-phase. The implementation is tailored to modern hardware features using vectorization and fast approximations of transcendental functions. A performance model and numerical examples verify the high degree of optimization.
We demonstrate the applicability and predictive power of the approach and the influence of scan strategy and geometry. Depending on the specific example, results can be obtained with moderate computational resources in a few hours to days. The numerical examples include a prediction of the microstructure on the full NIST AM Benchmark cantilever specimen. 
\end{abstract}

\keywords{powder bed fusion additive manufacturing, part-scale, microstructure, performance modeling}

\section{Introduction}

Laser powder bed fusion (LPBF) is a prominent additive manufacturing (AM) technique that allows the design and production of parts with complex geometry in a near-net-shape manner.
However, a successful build can require expensive trial-and-error runs beforehand or the adoption of overly conservative parameter choices.  The microstructure provides a critical insight into the material behavior of manufactured parts \cite{yang2016formation}. Understanding its evolution through numerical simulations offers the opportunity to significantly reduce the costs measured in time and money and enhance the physical understanding of the LPBF process. While subsequent heat treatment often changes the as-built microstructure, the information is highly relevant during processing, as significant differences in material behavior can lead to defects during processing and failure to build a part. Ultimately, a great promise lies in the local control of the microstructural phase composition as the specific application desires.

From a numerical modeling point of view, LPBF is a multi-scale and multi-physics problem \cite{meier2017thermophysical}. In this contribution, we focus on the effects on the microscale, specifically the composition and evolution of microstructure phases for the commonly used alloy Ti-6Al-4V.
The microstructure model in this contribution was first discussed in our contribution \cite{nitzler2021novel}. Based on the classification in \cite{furrer2009introduction}, it falls in the category of \textit{phenomenological} \cite{crespo2011modelling, lindgren2016simulation, murgau2012model, salsi2018modeling, zhang2019metallurgical} microstructure models, with alternative approaches being \textit{statistical} \cite{nie2014numerical, koepf2019numerical, rai2016coupled, ding2004microstructural, rolchigo2022exaca} and \textit{phase-field} \cite{chen2004quantitative, gong2015phase} models. A phenomenological microstructure model provides a reasonable trade-off between the costly evaluation of a phase-field model and the limited physical motivation of purely statistical models. A significant difference between \cite{nitzler2021novel} and other phenomenological models is that the evolution of the microstructural phases is governed by temperature-dependent diffusive and instantaneous forces driving the phase composition towards an equilibrium state. From a mathematical perspective, the model consists of coupled ordinary differential equations (ODEs) for the three phases, which are $\beta$-phase, stable $\alpha_s$-phase, and metastable martensitic $\alpha_m$-phase.

To the best of the authors' knowledge, no coupled thermo-microstructure simulation with scan-resolved tracks on the part-scale has been published before. Nevertheless, efforts in this direction were undertaken by some groups. In \cite{promoppatum2018numerical} and \cite{munk2022geometry}, the authors determined correlations between experimentally measured microstructural phase compositions and numerical results for the temperature field of larger parts without an explicit microstructure model. In \cite{berry2021toward}, the authors inform a phase-field model for microstructure evolution with a single-track melt pool simulation. The cellular automata microstructure model used in \cite{koepf2019numerical} aims at the part-scale by replicating the thermal information from a few representative layers and tracks over multiple layers. A similar strategy is used in \cite{turner2022exaam} and combined with a phase-field model for the sub-grain scale. Many authors integrate analytical Johnson-Mehl-Avrami-Kolmogorov (JMAK) equations into thermal or thermo-mechanical models \cite{crespo2011modelling, lindgren2016simulation, salsi2018modeling, zhang2019metallurgical}. In our publication \cite{nitzler2021novel}, we already determined the microstructure for application-motivated temperature profiles at selected points and for a quenching example of a large block. In contrast to all the cited references, this article presents a fully coupled thermo-microstructure model that considers the resolved laser scan track and is not restricted to regular geometries. The microstructure is determined in the whole domain at all points in time. 

An essential aspect of macroscale simulations is the question of the implementation's performance in terms of time to solution. For the coupled thermo-microstructure problem, the thermal history drives the evolution of the microstructure. Thus, a fast and accurate solution to the thermal problem is necessary.
In our previous work \cite{proell2023highly}, we presented a highly efficient solution to the thermal problem with a resolved laser scan track over hundreds of layers. In contrast to many existing approaches, we can simulate parts on the centimeter scale in a time frame on the order of hours or days.
The present contribution builds on that work and allows us to predict the composition of microstructure phases in the same setting with only marginally increased time to solution. To achieve this level of performance for the microstructure model, we carefully analyzed the many conditional branches in the governing equations. Our implementation is tailored to modern hardware capabilities as it can utilize vectorization efficiently and considers the need to reduce memory transfer as much as possible. Efficient approximations of transcendental functions \cite{schraudolph1999fast, malossi2015fast, perini2018fast}, e.g., the exponential function, which can be vectorized efficiently, are discussed. We present a detailed performance analysis with the help of a roofline performance model. 

The article is structured as follows: after briefly reviewing the thermal and microstructure model, we focus on the latter's numerical discretization and implementation details. Specifically, we discuss the implementation tailored to modern hardware and present efficient approximations for expensive transcendental functions. We study the implementation performance on benchmarks and application examples. The investigated, practically relevant examples demonstrate a wide range of applicability of the approach and fast solution times. A notable example is the NIST AMBench 2022 cantilever geometry \cite{AMbench2022}, for which we predict the as-built microstructure.

\section{Coupled thermo-microstructure model}

This section summarizes model equations. An emphasis is placed on details especially relevant to the efficient solution strategy presented later in this article. We refer to the respective publications for the full details of the thermal \cite{proell2023highly} and microstructure \cite{nitzler2021novel} models.

\subsection{Thermal model}
First, we briefly summarize the thermal part of the problem following \cite{proell2023highly}. The temperature field $T$  is determined in the domain $\Omega$ by solving the heat equation:
\begin{align}
\label{eq:heat_equation}
\rho c \pfrac{T}{t} &=  - \nabla \cdot {\vec{q}} + q_\text{vol} ,\quad \vec{q} = -k(T) \nabla T && \text{in } \Omega,
\end{align}
Here, $\rho$ is the density and 
$c$ is the specific heat capacity of the material. The temperature and state-dependent heat conductivity $k$ can be computed from the liquid fraction $g(T)$, defined as
\begin{align}
\label{eq:liquid_fraction}
g(T) = \begin{cases}
0, & T < T_s,\\
\frac{T-T_s}{T_l-T_s}, &T_s \leq T \leq T_l\\
1, &T > T_l,
\end{cases}
\end{align}
where $T_s$ and $T_l$ are the solidus and liquidus temperature. The time-dependent consolidated fraction
\begin{align}
\label{eq:consolidated_fraction}
r_c(t) = \begin{cases}
1, & \text{if } r_c(0)=1 \text{ (i.e. initially consolidated)}\\
\underset{\tilde{t}<t}{\max}\, g(T(\tilde{t})), & \text{if } r_c(0)=0 \text{ (i.e. initially powder)}\\
\end{cases}.
\end{align}
captures the irreversible powder-to-melt transition and allows setting the initial material state.
From \eqref{eq:liquid_fraction} and \eqref{eq:consolidated_fraction}, the actual fractions of powder ($p$), melt ($m$) and solid ($s$) material are computed as
\begin{align}
\label{eq:material_state_fractions}
r_p(r_c) = 1 - r_c,\quad
r_m(T) = g(T),\quad
r_s(T,r_c) = r_c -g(T),
\end{align}
and finally, the temperature- and history-dependent heat conductivity $k(T,r_c)$ is found:
\begin{align}
\label{eq:material_parameter_interp}
k(T,r_c) = r_p(r_c) k_p + r_m(T) k_m + r_s(T,r_c)  k_s,
\end{align}
where $k_p$, $k_s$ and $k_m$ are the parameters for a single state. 

A cylindrical volumetric heat source $q_\text{vol}$  formulated in a local coordinate system $(\hat{x}, \hat{y}, \hat{z})$ models the incident energy from a moving laser beam:
\begin{align}
\label{eq:heat_soruce_cylindrical}
q_\text{vol} = \begin{cases} \frac{2 W_\text{eff}}{\pi R^2h_\text{powder}}\exp\left(\frac{-2(\hat{x}^2+\hat{y}^2)}{R^2}\right), & \text{if } 0< \hat{z} < -h_\text{powder} \\
0, &\text{otherwise}
\end{cases},
\end{align}
Here, $R$ is the effective beam radius of the incident energy beam, $W_\text{eff}$ is the effective  power and $h_\text{powder}$ is the powder layer thickness.

The necessary initial and boundary conditions  for the heat equation \eqref{eq:heat_equation} are given as:
\begin{align}
    \label{eq:init_condition}
T &= T_0 && \text{in } \Omega \text{ for } t=0,\\
\label{eq:bc_dirichlet}
T &= T_0 && \text{on } \Gamma_D,\\
\label{eq:bc_neumann}
\vec{q} \cdot \vec{n} &= 0 && \text{on } \Gamma_N,\\
\label{eq:bc_rad_evap}
\vec{q} \cdot \vec{n} &= q_\text{rad} + q_\text{evap} && \text{on } \Gamma_{RE},\\
\label{eq:bc_radiation}
& q_\text{rad} =\epsilon \sigma_S(T^4-T_\infty^4), \\
\label{eq:bc_evaporation}
& q_\text{evap} = 0.82 C_P\exp\left[ -C_T \left(\frac{1}{[T]} -\frac{1}{T_v}\right)\right] \sqrt{\frac{C_M}{[T]}} \ (h_v +c([T]-T_{h,0})),\ \text{if } [T] > T_v.
\end{align}

All material is initially pre-heated to a temperature $T_0$.
The material parameters required for the initial boundary value problem are listed in Table~\ref{tab:thermal_material_parameters}. In addition, 
$\sigma_S$ in \eqref{eq:bc_radiation} is the Stefan-Boltzmann constant. To avoid numerical challenges arising from the strong nonlinearity in the evaporation term \eqref{eq:bc_evaporation}, the temperature $[T]$ is limited to a maximum value $T_\text{max} > T_v$. In this study, we opt for $T_\text{max} = T_v + 1000\, \si{\kelvin}$, which ensures numerical stability without affecting the overall results.

\begin{table}[]
    \centering
    \caption{Thermal model parameters for Ti-6Al-4V.}
    \label{tab:thermal_material_parameters}
    \begin{tabular}{llrl}
			\toprule
			Symbol & Property & Value & Unit\\
			\midrule
			$k_{ms}$ & Thermal conductivity in melt and solid phase & 28.6 & $\si{\watt\per\metre\per\kelvin}$ \\
			$k_{p}$ & Thermal conductivity in powder phase & 0.286 & $\si{\watt\per\metre\per\kelvin}$ \\
			$\rho$ & Density & 4090 & $\si{\kilogram\per\cubic\meter}$\\
			$c$ & Specific heat capacity & 1130 & $\si{\joule\per\kilogram\per\kelvin}$ \\
			$T_s$ & Solidus temperature & 1878 & $\si{\kelvin}$\\
			$T_l$ & Liquidus temperature & 1928 & $\si{\kelvin}$\\
			$T_\infty$ &Ambient temperature & 293 & $\si{\kelvin}$\\
			$\epsilon$ & Emissivity & 0.7 & --\\
			\midrule
			$T_v$ & Boiling temperature & 3130 & $\si{\kelvin}$ \\
			$C_P$ & Recoil pressure factor & 54 & $\si{\kilo\pascal}$ \\
			$C_T$ & Recoil pressure temperature factor & \num{5.07e4} & $\si{\kelvin}$ \\
			$C_M$ & Heat loss temperature factor & \num{9.15e-4} & $\si{\kelvin\square\second\per\square\meter}$ \\
		 $M$ &  Molar mass &  0.0478 &  $\si{\kilogram\per\mole}$ \\
			$h_v$ & Latent heat of evaporation & 8.84 & $\si{\mega\joule\per\kilogram}$ \\
			$T_{h,0}$ & Enthalpy reference temperature & 538 & $\si{\kelvin}$ \\
			\bottomrule
         & 
    \end{tabular}
\end{table}

\subsection{Microstructure model}

A phenomenological model for the microstructure evolution was presented in \cite{nitzler2021novel}.
This model focuses on the three most important phases\footnote{Note that the word `phase' refers to the microstructure phases $\alpha_s$, $\alpha_m$, and $\beta$. In contrast, when we distinguish material into powder, melt, and solid, we speak of the `state' of the material.}, $\beta$, $\alpha_s$ and $\alpha_m$. While the thermal model needs to consider the powder material state, we can neglect it in the context of the microstructure model. We define phase fractions $X_i \in [0;1]$ for the microstructure phases along with elementary continuity constraints:
\begin{align}
    \label{eq:phase_continuity_sol_liq}
    X_\text{sol}+X_\text{liq} &= 1,\\
    \label{eq:phase_continuity_alpha_beta}
    X_{\alpha}+X_{\beta} &= X_\text{sol},\\
    \label{eq:phase_continuity_alpha_s_m}
    X_{\alpha_s} + X_{\alpha_m} &= X_\alpha.
\end{align}
The liquid phase fraction $X_\text{liq}$ and, by virtue of \eqref{eq:phase_continuity_sol_liq}, the solid phase fraction $X_\text{sol}$ are again computed according to $g(T)$ \eqref{eq:liquid_fraction} as
\begin{align}
    X_\text{liq} = g(T), \quad X_\text{sol} = 1 - g(T).
\end{align}
While, for now, the definitions of $X_\text{sol}$ and $X_\text{liq}$ seem redundant to the state described in \eqref{eq:material_state_fractions}, they make for a cleaner notation, and the duplicate definition allows us to treat them slightly differently in terms of numerics later on. Also, note that the powder state is now implicitly part of the solid state. The transformation from solid to liquid state and vice versa is thus complete, and the remainder of this section focuses on the microstructure phase transformations.

Before the evolution equations for the phase fractions can be defined, we introduce two equilibrium phase fractions associated with material cooling from a molten state to ambient temperature $T_\infty$. All newly solidified solid material consists entirely of $\beta$-phase. On further cooling, the $\beta$-phase can transform into stable $\alpha_s$-phase. 
The fraction $X_\alpha^\text{eq}(T)$ determines the amount of stable $\alpha$-phase at a given temperature in thermodynamic equilibrium. It follows an exponential Koistinen-Marburger law:
\begin{align}
\label{eq:equilibrium_phase_as}
    X_\alpha^\text{eq}(T) = \begin{cases}
        0.9 & \text{for } T < T_{\alpha_s,\text{end}},\\
        1 - \exp\lbrack -k_\alpha^\text{eq}(T_{\alpha_s, \text{start}} - T) \rbrack & \text{for } T_{\alpha_s,\text{end}} \leq T \leq T_{\alpha_s,\text{start}},\\
        0 & \text{for } T > T_{\alpha_s,\text{start}},
    \end{cases}
\end{align}
where the parameters $k_\alpha^\text{eq}$, $T_{\alpha_s,\text{start}}$ and $T_{\alpha_s,\text{end}}$ are obtained either directly from the literature or fitted to experimental data in \cite{nitzler2021novel}. Their values are listed in Table~\ref{tab:micro_parameters}.
Notably, the actual fraction of stable $\alpha$-phase $X_{\alpha_s}$ at a given temperature is, in general, not identical to the equilibrium value in \eqref{eq:equilibrium_phase_as}, which is only reached when the cooling rate is very low. Instead, the equilibrium value can be interpreted as the long-term solution for $t\rightarrow\infty$. The formation of $\alpha_s$ is driven by a diffusion process as detailed below. If the cooling rate is high, this diffusion process is inhibited, and metastable martensite $\alpha_m$-phase forms instead. Again, we define a pseudo equilibrium phase fraction $X_{\alpha_m,0}^\text{eq}$ for martensitic $\alpha_m$-phase for high cooling rates:
\begin{align}
\label{eq:equilibrium_phase_am0}
    X_{\alpha_m,0}^\text{eq}(T) = \begin{cases}
        0.9 & \text{for } T < T_{\infty}\\
        1 - \exp\lbrack -k_{\alpha_m}^\text{eq}(T_{\alpha_m, \text{start}} - T) \rbrack & \text{for } T_{\infty} \leq T \leq T_{\alpha_m,\text{start}}\\
        0 & \text{for } T > T_{\alpha_s,\text{start}}
    \end{cases}
\end{align}
with the parameters $k_{\alpha_m}^\text{eq}$ and $T_{\alpha_m,\text{start}}$ listed in  Table~\ref{tab:micro_parameters}. Again, the actual fraction of martensitic $\alpha_m$-phase $X_{\alpha_m}$ at a given temperature is not necessarily identical to the equilibrium value $ X_{\alpha_m,0}^\text{eq}$. 
Equations \eqref{eq:equilibrium_phase_as} and \eqref{eq:equilibrium_phase_am0} for the equilibrium fractions are incompatible for specific temperature values, in the sense that, e.g., for $T=T_\infty$ their values sum to 1.8, which is larger than 1. The transformation from $\beta$-phase to $\alpha_s$-phase happens in a strictly higher temperature interval than the transformation from  $\beta$-phase to $\alpha_m$-phase for the material parameters listed in Table~\ref{tab:micro_parameters}. For this reason, the incompatibility is resolved by the introduction of the following effective pseudo-equilibrium phase fraction:
\begin{align}
\label{eq:equilibrium_phase_am}
X_{\alpha_m}^\text{eq}(T) = X_{\alpha_m,0}^\text{eq}(T)\cdot\frac{0.9 - X_{\alpha_s}}{0.9}.
\end{align}
This means that the martensite equilibrium fraction $X_{\alpha_m}^\text{eq}$ is corrected based on the already formed fraction of stable $\alpha_s$-phase, with the effect that $X_{\alpha_s} +X_{\alpha_m}^\text{eq} \leq 0.9 $. 
\begin{figure}
    \centering
    \includegraphics[width=.4\linewidth]{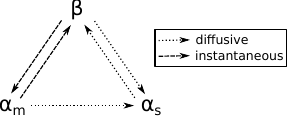}
    \caption{Possible transformation paths between different microstructure phases.}
    \label{fig:phase_transformations_overview}
\end{figure}
The formation and dissolution of the three phases $\alpha_s$, $\alpha_m$ and $\beta$ is governed by the following ordinary differential equations:
\begin{align}
    \label{eq:alpha_s_total_time_deriv}
    \dot{X}_{\alpha_s} &= \dot{X}_{\beta\rightarrow\alpha_s} + \dot{X}_{\alpha_m\rightarrow\alpha_s} - \dot{X}_{\alpha_s\rightarrow\beta},\\
    \label{eq:alpha_m_total_time_deriv}
    \dot{X}_{\alpha_m} &= \dot{X}_{\beta\rightarrow\alpha_m} - \dot{X}_{\alpha_m\rightarrow\alpha_s} - \dot{X}_{\alpha_m\rightarrow\beta},\\
    \label{eq:beta_total_time_deriv}
    \dot{X}_{\beta} &= \dot{X}_{\alpha_m\rightarrow\beta} +\dot{X}_{\alpha_s\rightarrow\beta} -\dot{X}_{\beta\rightarrow\alpha_m} -\dot{X}_{\beta\rightarrow\alpha_s},
\end{align}
where $\dot{X}_{i\rightarrow j}$ is the formation of phase $j$ from phase $i$ or, equivalently, the dissolution from phase $i$ into phase $j$. Note that $\dot{X}_{i\rightarrow j} \neq \dot{X}_{j\rightarrow i}$, since the reverse transformation might not even exist (e.g., for the transformation from martensite to stable $\alpha_s$ phase) or is governed by different kinetics. A graphical overview of the transformation processes is shown in Figure~\ref{fig:phase_transformations_overview}. The transformation $\alpha_m \leftrightarrow \beta$  occurs on a much shorter time scale than $\alpha_s \leftrightarrow \beta$. Therefore, we treat the phase change $\alpha_m \leftrightarrow \beta$ as instantaneous, while the transformations $\alpha_s \leftrightarrow \beta$ and $\alpha_m \rightarrow \alpha_s$ are modeled as a diffusive process.

For $T < T_s$, the sum of \eqref{eq:alpha_s_total_time_deriv}--\eqref{eq:beta_total_time_deriv} yields
\begin{align}
 \dot{X}_{\alpha_s} + \dot{X}_{\alpha_m} + \dot{X}_{\beta} = 0 \quad \text{if}\quad T < T_s,
\end{align}
allowing us to directly express the $\beta$-phase fractions as
\begin{align}
\label{eq:beta_post_processing}
    X_\beta = X_\text{sol} - X_{\alpha_s}- X_{\alpha_m}.
\end{align}
This makes integration of \eqref{eq:beta_total_time_deriv} unnecessary.

The following set of logistic differential equations describes the diffusion-driven transformation processes:
\begin{align}
    \label{eq:beta_to_alpha_s}
    \dot{X}_{\beta\rightarrow\alpha_s} &= \begin{cases}
        k_{\alpha_s}(T) \left(X_{\alpha_s}\right)^{\frac{c_{\alpha_s} - 1}{c_{\alpha_s}}}\left(X_\alpha^\text{eq} - X_\alpha\right)^{\frac{c_{\alpha_s} + 1}{c_{\alpha_s}}} &\text{if }X_\alpha < X_\alpha^\text{eq},\\
        0 &\text{otherwise}
    \end{cases}\\
    \label{eq:alpha_m_to_alpha_s}
    \dot{X}_{\alpha_m\rightarrow\alpha_s} &= \begin{cases}
        k_{\alpha_s}(T) \left(X_{\alpha_s}\right)^{\frac{c_{\alpha_s} - 1}{c_{\alpha_s}}}\left(X_{\alpha_m}\right)^{\frac{c_{\alpha_s} + 1}{c_{\alpha_s}}} &\text{if }X_{\alpha_m} > 0,\\
        0 &\text{otherwise}
    \end{cases}\\
    \label{eq:alpha_s_to_beta}
    \dot{X}_{\alpha_s\rightarrow\beta} &= \begin{cases}
        k_{\beta}(T) \left(0.9 - X_{\alpha}\right)^{\frac{c_{\beta} - 1}{c_{\beta}}}\left(X_{\alpha} - X_{\alpha}^\text{eq}\right)^{\frac{c_{\beta} + 1}{c_{\beta}}} &\text{if }X_{\alpha} > X_\alpha^\text{eq},\\
        0 &\text{otherwise}
    \end{cases}
\end{align}
which are completed by temperature-dependent diffusion rates,
\begin{align}
    k_{\alpha_s}(T) = \frac{k_1}{1+\exp[-k_3(T-k_2)]} \quad \text{and} \quad k_{\beta}(T) = f \cdot k_{\alpha_s}(T).
\end{align}
In total, six parameters, $c_{\alpha_s}$, $c_\beta$, $k_1$, $k_2$, $k_3$ and $f$, govern the diffusion process. Their values were determined by fitting experimental data in \cite{nitzler2021novel} and are summarized in Table~\ref{tab:micro_parameters_diffusion}. 
When cooling down below $T_{\alpha_m,\text{start}}$ the martensite phase fraction instantaneously follows the pseudo-equilibrium phase fraction:
\begin{align}
    \label{eq:inst_martensite_formation}
    X_{\alpha_m} = X_{\alpha_m}^\text{eq}, \quad \text{if} \quad T < T_{\alpha_m,\text{start}}
\end{align}
which is already corrected according to \eqref{eq:equilibrium_phase_am} such that martensite only forms from $\beta$-phase if the $\alpha_s$-phase is below the equilibrium concentration. When heating material, the martensite phase fraction instantaneously 
dissolves into $\beta$-phase according to:
\begin{align}
    \label{eq:inst_martensite_dissolution}
    X_{\alpha_m} = \max(X_{\alpha}^\text{eq} - X_{\alpha_s}, 0), \quad \text{if} \quad X_{\alpha} > X_{\alpha}^\text{eq}
\end{align}
This implies that no previously formed martensite remains when $T > T_{\alpha_s, \text{start}}$. To fully dissolve the stable $\alpha_s$-phase when further heating the material, we introduce a regularization,
\begin{align}
    X_{\alpha_s} = \min\left(X_{\alpha_s}, 0.9\frac{T_{\alpha_s, \text{start}} + T_{\alpha_s, \text{reg}} - T}{ T_{\alpha_s, \text{reg}}}\right), \quad \text{if} \quad T_{\alpha_s, \text{start}} < T < T_{\alpha_s, \text{reg}},
\end{align}
Thus, the $\alpha_s$-phase fraction decreases linearly from a maximum of 0.9 to zero in the regularization interval $[ T_{\alpha_s, \text{start}},  T_{\alpha_s, \text{start}}  + T_{\alpha_s, \text{reg}}]$. In this contribution, we choose $T_{\alpha_s, \text{reg}} = 100\, \si{\kelvin}$.
The regularization has the effect that no $\alpha_s$-phase but only $\beta$-phase remains once the solidus temperature $T_\text{sol}$ is reached and material begins to melt.
Without the regularization, it would be possible to retain $\alpha_s$-phase until $T_\text{sol}$; however, in that case, one would need to model the melting of a mixture of two phases.
Since the exact phase composition in the high-temperature region is not of interest and due to the lack of experimental data, we choose the outlined strategy for temperatures above $T_{\alpha_s, \text{start}}$.

\begin{table}
\centering
\caption{Parameters of the microstructure evolution model. A detailed analysis and literature review of all parameters is given in \cite{nitzler2021novel}.}
\label{tab:micro_parameters}
\begin{tabular}{llll}
     \toprule
     Parameter & Description & Value & Unit\\
     \midrule
     $T_l$ & Liquidus temperature & 1928 &\si{\kelvin}\\
     $T_s$ & Solidus temperature & 1878 &\si{\kelvin}\\
     $T_{\alpha_s,\text{start}}$ & Upper end of temperature range for $\beta\rightarrow \alpha_s$ transformation & 1273 &\si{\kelvin} \\
     $T_{\alpha_s,\text{end}}$ &  Lower end of temperature range for $\beta\rightarrow \alpha_s$ transformation & 935 &\si{\kelvin}\\
     $T_{\alpha_m,\text{start}}$ &  Upper end of temperature range for $\beta\rightarrow \alpha_m$ transformation & 848 &\si{\kelvin}\\
     $T_\infty$ & Ambient temperature & 293 &\si{\kelvin} \\
     $k_\alpha^\text{eq}$ & $\alpha_s$-phase equilibrium concentration constant & 0.0068 &\si{\per\kelvin} \\
     $k_{\alpha_m}^\text{eq}$ & $\alpha_m$-phase equilibrium concentration constant & 0.00415 &\si{\per\kelvin}\\
     \bottomrule
\end{tabular}
\end{table}

\begin{table}
\centering
\caption{Fitted parameters of the microstructure evolution model. A detailed explanation of the calibration process is given in \cite{nitzler2021novel}.}
\label{tab:micro_parameters_diffusion}
\begin{tabular}{lll}
     \toprule
     Parameter & Value & Unit\\
     \midrule
     $c_{\alpha_s}$ & 2.51 & --\\
     $c_{\beta}$ & 11.0 & --\\
     $k_1$ & 0.294 & \si{\per\second}\\
     $k_2$ & 850 &\si{\kelvin}\\
     $k_3$ & 0.0337 &\si{\per\kelvin}\\
     $f$ & 3.8& --\\
     \bottomrule
\end{tabular}
\end{table}

\section{Numerical methods and efficient implementation}

The thermal problem is solved with an efficient FEM implementation \cite{proell2023highly} based on fast operator evaluation \cite{kronbichler2012generic}. Implementation is performed with the \texttt{deal.II} library \cite{arndt2023deal}. Notably, we use an explicit scheme for the active laser phase where small time step sizes are necessary to obtain a continuous melt track. In the interlayer cool down phase after every layer, we use an implicit scheme allowing larger time step sizes. This approach is extended to the microstructure model where we introduce a fast explicit and a more accurate and robust implicit scheme.

The microstructure model is integrated into the existing approach as a one-way coupled problem that is solved after the thermal model in every time step. Since the microstructure model does not explicitly depend on the spatial coordinate, the problem is fully decoupled in space. The ODEs can be solved independently at every point in space. It is sufficient to place the degrees of freedom (DoFs) of the microstructure model (phases $X_\beta$, $X_{\alpha_s}$ and $X_{\alpha_m}$) in the same spatial positions as the thermal DoFs (temperature $T$). This choice has the advantage that no communication and interpolation is necessary to obtain the temperature to evaluate the microstructure model.

The spatial discretization is realized as an adaptive mesh. When activating a new layer, more refined cells are placed in the highest active layer and a few layers beneath. The mesh can be coarsened at later stages when the material history allows for it. Previously, the only history variable that needed to be considered for the thermal model was the consolidated fraction $r_c$. In this contribution, the three microstructure phases also represent a material history that should not be coarsened inadvertently. Therefore, an octant consisting of eight sibling cells of equal refinement level will only be coarsened when, for every microstructure phase, all its values are sufficiently close to each other.
For more general details of the adaptive and growing mesh, the reader is referred to \cite{proell2023highly}.

\subsection{Time integration of microstructure model}
To ease the notation, the unknown fractions of stable and martensitic $\alpha$-phase are collected in a state vector $\vec{m} = [X_{\alpha_s}, X_{\alpha_m}]$. The $\beta$-phase fractions can be processed for a given state and temperature as,
\begin{align}
\label{eq:time_integration_beta_postprocess}
X_\beta(\vec{m}, T) = X_\text{sol}(T) - X_{\alpha_s} - X_{\alpha_m}.
\end{align}
The right-hand side of the differential equations \eqref{eq:alpha_s_total_time_deriv}--\eqref{eq:alpha_m_total_time_deriv} is split into a diffusion-based and instantaneous contribution:
\begin{align}
\label{eq:diff_inst_split}
    \dot{\vec{m}} = \begin{bmatrix}
        \dot{X}_{\alpha_s}\\
       \dot{X}_{\alpha_m}
    \end{bmatrix} = \underbrace{\begin{bmatrix}
        \dot{X}_{\beta\rightarrow\alpha_s} + \dot{X}_{\alpha_m\rightarrow\alpha_s} - \dot{X}_{\alpha_s\rightarrow\beta}\\
       -\dot{X}_{\alpha_m\rightarrow\alpha_s}
    \end{bmatrix}}_{=: \dot{\vec{m}}_\text{diff}} + \underbrace{\begin{bmatrix}
        0\\
       \dot{X}_{\beta\rightarrow\alpha_m} -\dot{X}_{\alpha_m\rightarrow\beta}
    \end{bmatrix}}_{=: \dot{\vec{m}}_\text{inst}}
\end{align}
Integrating \eqref{eq:diff_inst_split} from a point in time $t^n$ to a point in time $t^{n+1} = t^n + \Delta t$ yields,
\begin{align}
\label{eq:time_integration_general}
    \vec{m}^{n+1} &= \vec{m}^n +  \int_{t^n}^{t^{n+1}} \dot{\vec{m}}_\text{diff} \, \dd t + \int_{t^n}^{t^{n+1}} \dot{\vec{m}}_\text{inst} \, \dd t
\end{align}
where the superscript indicates at which point in time a quantity is evaluated. The phase fractions $\vec{m}^n$ and the temperature $T^{n+1}$ are known. A numerical time integration scheme will approximate the first integral over the diffusion-driven contribution. Note that the second integral over the (non-differentiable) instantaneous change rate yields a finite value for the absolute change. These instantaneous changes are defined in \eqref{eq:inst_martensite_formation} and \eqref{eq:inst_martensite_dissolution}.

\paragraph{Explicit time integration}
We split \eqref{eq:time_integration_general} into a two-stage process, where integration of the diffusion-based term is performed with a forward Euler scheme towards an intermediate state $\tilde{\vec{m}}^{n+1}$, followed by a correction step:
\begin{align}
    \label{eq:explicit_integration}
    \tilde{\vec{m}}^{n+1} 
    & = \vec{m}^n + \Delta t\,\dot{\vec{m}}_\text{diff}(T^{n}, \vec{m}^{n}),\\
    \label{eq:explicit_correction}
    {\vec{m}}^{n+1} &= \vec{g}(T^{n+1}, \tilde{\vec{m}}^{n+1}).
\end{align}
Here, we define a correction function $\vec{g}(T, \vec{m})$, which contains the instantaneous changes for the martensite phase as well as corrections that are necessary to satisfy the continuity constraints \eqref{eq:phase_continuity_sol_liq}--\eqref{eq:phase_continuity_alpha_s_m}. Furthermore, if either $X_{\alpha_s}$ or $X_{\alpha_m}$ falls below zero, it is instead set to zero. If $X_\alpha$ exceeds the maximum equilibrium fraction of 0.9, both, $X_{\alpha_s}$ or $X_{\alpha_m}$ are reduced while maintaining the ratio  $X_{\alpha_s}/X_{\alpha_m}$.

Two exceptional cases become apparent when examining \eqref{eq:beta_to_alpha_s} and \eqref{eq:alpha_s_to_beta}. Both equations pose a problem for the explicit time integration scheme presented so far. Evaluating \eqref{eq:beta_to_alpha_s} for $X_{\alpha_s} = 0, X_{\alpha_m} = 0, X_\beta = 1.0$ gives a rate of zero, which implies that a solution computed with the explicit scheme can never evolve out of the initial state. Thus, we initiate the diffusion process with the help of an approximate analytical solution described in more detail in \cite{nitzler2021novel}. The same strategy can be applied to \eqref{eq:alpha_s_to_beta} which suffers from the same problem for $X_{\alpha_s} = 0.9, X_{\alpha_m} = 0, X_\beta = 0.1$.
Due to the numeric values of the physical constants, the second case is prone to cancellation of significant digits. An appropriate reformulation can be found in Appendix~\ref{appendix:analytical_solution}.

The presented explicit time integration scheme does not come with a stability limit due to the correction function $\vec{g}(T, \vec{m})$ being applied after every step. Therefore, the solution cannot grow arbitrarily large, and classical stability considerations do not apply to the specific scheme used here. Still, the time step size should be chosen within limits to obtain a sufficiently accurate solution.
\paragraph{Implicit time integration}
We use a Crank-Nicolson time integration scheme for the diffusion-based term and reuse the same correction function $\vec{g}$ as in the explicit case:
\begin{align}
    \label{eq:implicit_integration}
    \tilde{\vec{m}}^{n+1} 
    & = \vec{m}^n + \frac{1}{2}\Delta t\left(\dot{\vec{m}}_\text{diff}(T^{n+1}, \vec{m}^{n+1}) + \dot{\vec{m}}_\text{diff}(T^{n}, \vec{m}^{n})\right),\\
    \label{eq:implicit_correction}
    {\vec{m}}^{n+1} &= \vec{g}(T^{n+1}, \tilde{\vec{m}}^{n+1}).
\end{align}
The nonlinear equation \eqref{eq:implicit_integration} is solved employing a fixed-point iteration:
\begin{align}
\label{eq:implicit_fp_iteration}
    \vec{m}_{i+1}^{n+1} &= \vec{g}\left(\vphantom{\frac{1}{2}}\right.T^{n+1}, \underbrace{\vec{m}^n + \frac{1}{2}\Delta t\left(\dot{\vec{m}}_\text{diff}(T^{n+1}, \vec{m}_i^{n+1}) + \dot{\vec{m}}_\text{diff}(T^{n}, \vec{m}^{n})\right)}_{\tilde{\vec{m}}_{i+1}^{n+1}} \left.\vphantom{\frac{1}{2}}\right),\\
    \vec{m}_{0}^{n+1} &= \vec{m}^n,
\end{align}
until the weighted root-mean-square (WRMS) norm of the residual%
\begin{align}
     \vec{r}^{n+1}_{i+1}  &=   \vec{m}_{i+1}^{n+1} - \left(\vec{m}^n + \frac{1}{2}\Delta t\left(\dot{\vec{m}}_\text{diff}(T^{n+1}, \vec{m}_{i+1}^{n+1}) + \dot{\vec{m}}_\text{diff}(T^{n}, \vec{m}^{n})\right)\right) \\
    &=  \frac{\Delta t}{2} \left(\dot{\vec{m}}_\text{diff}(T^{n+1}, \vec{m}_{i}^{n+1}) - \dot{\vec{m}}_\text{diff}(T^{n+1}, \vec{m}_{i+1}^{n+1})\right) 
\end{align}
falls below a threshold. The WRMS norm is defined as
\begin{align}
    \vert\vert \vec{a} \vert\vert_\text{WRMS} = \frac{1}{N} \sum_{j=1}^N (a_jw_j)^2, \quad \text{where } w_j = (\varepsilon_\text{abs} + \bar{a}_j\varepsilon_\text{rel})^{-1}
\end{align}
with an absolute tolerance $\varepsilon_\text{abs} = \num{1e-10}$ and a relative tolerance $\varepsilon_\text{rel} = \num{1e-3}$.
It can easily be verified that the iteration scheme prescribed in \eqref{eq:implicit_fp_iteration} is a contraction on $X_i \in(0,0.9)$, and thus converges to a unique solution if the time step size $\Delta t$ is sufficiently small. This proof holds for time step sizes up to around $\num{0.001}\,\si{\second}$. However, we experimentally observe fast and robust convergence for time step sizes up to $\num{1}\,\si{\second}$. We use a subcycling scheme to achieve sufficient accuracy, where a large time step performed in the thermal model is subdivided into substeps not exceeding a maximum size $\Delta t_\text{sub} = 0.01\, \si{\second}$.

\subsection{Vectorized computation}

Modern CPUs support vector operations on specialized execution units that perform the same operation on a (small) array of different data, an idiom commonly called \textit{Single Instruction Multiple Data} (SIMD). This small array has $n_\text{lanes}$ entries and will be referred to as a \textit{vectorized array}. A useful C++ type \texttt{VectorizedArray} overloading basic arithmetic operations is provided by the \texttt{deal.II} library. Similar data types are available in the experimental C++ standard implementation \cite{stdsimd} or as stand-alone libraries \cite{vectorclass}. The overloaded operations simultaneously perform the arithmetic operation on all lanes by calling the respective, hardware-specific intrinsic functions.
For instance, the latest Intel AVX512 instruction set architecture supports eight concurrent double-precision operations. This capability promises to significantly speed up computation-heavy code paths by a factor of $n_\text{lanes}$ when fully utilized. Using SIMD operations demands contiguous data storage in memory for maximum performance benefits. Furthermore, the concurrently processed data should be independent, i.e., the computation in one vector lane may not depend on a computation in another lane of the same vector.

\begin{figure}
    \centering
    \includegraphics[width=\linewidth]{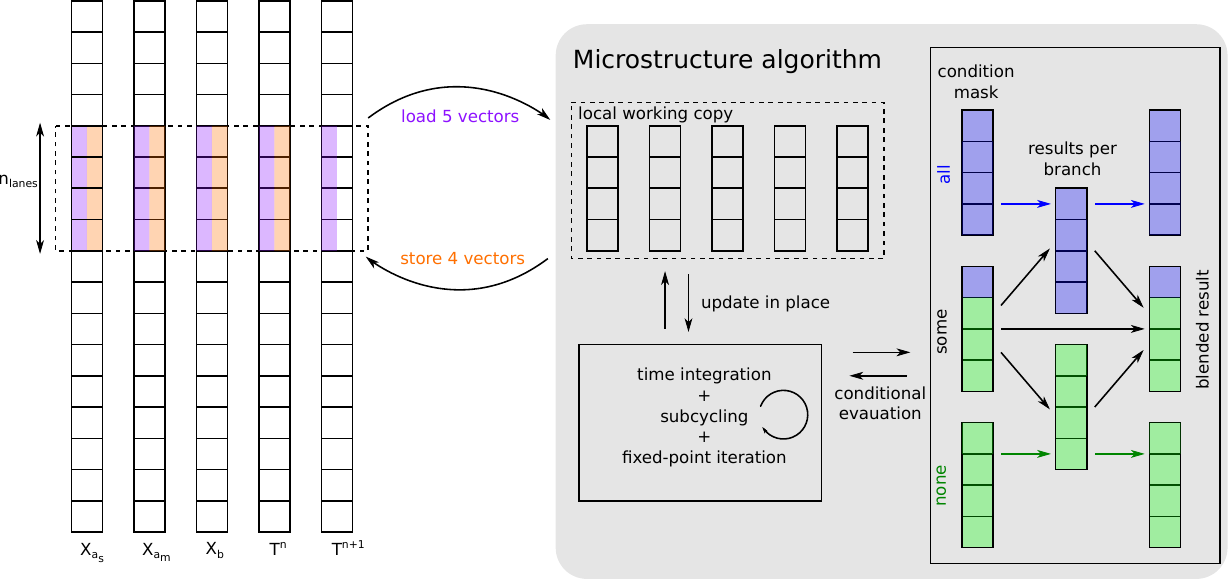}
    \caption{Illustration of vectorized microstructure algorithm. The algorithm works simultaneously on a local working copy of $n_\text{lanes}$ entries of the five global data vectors. Time integration, fixed-point iteration, and (optional) subcycling are all performed locally. All conditional computations are performed by blending the results of different conditional branches based on the condition mask.}
    \label{fig:vectorization_load_store_layout}
\end{figure}

The classical and automatic approach to vectorization is to look at inner loops or plain code and let the compiler identify nearby operations of the same kind that could go to different lanes. The results of auto-vectorization are poor if the loop kernel contains code beyond basic arithmetic, e.g., conditional branches and transcendental function calls. Therefore, we choose to vectorize the outer loop over all points in the mesh in batches of size $n_\text{lanes}$ and then solve the microstructure ODEs on a batch of points stored in the different lanes. The different quantities on a batch of points are loaded into the \texttt{VectorizedArray} data structure. This approach results in an array-of-struct-of-array layout, with the inner array being a \texttt{VectorizedArray}, which allows a concise implementation without explicitly writing the inner-most vectorized loop.
However, the numerous conditional branches in the microstructure model do not always allow the same operation to be performed on all lanes.
From an implementation standpoint, three distinct scenarios for the conditional
branches of equations can be distinguished:

\begin{enumerate}
\item \textit{All} vector lanes require evaluation of an expression. In this case, the expression is evaluated 
for all vector lanes using intrinsic functions.
\item \textit{None} of the vector lanes require evaluation of an expression. In this case, the expression is not evaluated.
\item \textit{Some} but not all of the vector lanes require the evaluation of an expression. In this case, the expression is evaluated for \textit{all} vector lanes, but the result is only stored in \textit{some} vector lanes that require it. Note that the unused additional computations do often not impact the evaluation time compared to an unvectorized implementation.\footnote{Note that wider vectors might lead to slightly lower clock frequencies on some hardware and complicated functions, like divisions or square roots, might have lower throughput when executed on wider vectors; nonetheless, the overwhelming share of operations has the same throughput for 1 or $n_\text{lanes}$ results.}
\end{enumerate}

A condition mask is computed for every branch to see which of the three scenarios is active. In the third case, the results of different branches are combined by blending the results with the condition mask. The general strategy is also visualized in Figure~\ref{fig:vectorization_load_store_layout}.

The strategy outlined above only makes sense when an efficient vectorized implementation of every required (mathematical) function is available. Although it is possible to fall back to compute expressions on the vector lanes sequentially, doing this in the scenario of mixed operations can, in the worst case, lead to an \textit{increase} in computation time on the order of $\mathcal{O}\left(n_\text{lanes}\right)$ compared to an unvectorized implementation\footnote{If every lane requires a different kind of operation, $\mathcal{O}\left(n_\text{lanes}\right)$ operations are necessary in the unvectorized case. However, when the computation is vectorized and every kind of operation is unnecessarily performed on all lanes,  $\mathcal{O}\left(n_\text{lanes}^2\right)$ operations are necessary.}. 

\begin{algorithm}
\caption{Time integration of microstructure model with vectorized data.}\label{alg:timeint}
\DontPrintSemicolon
\SetAlgoLined
\KwData{$\Delta t$, Global vectors $X_{\alpha_s}^n$, $X_{\alpha_m}^n$, $X_{\beta}^n$, $T^n$, $T^{n+1}$ }
\KwResult{Global vectors $X_{\alpha_s}^{n+1}$, $X_{\alpha_m}^{n+1}$, $X_{\beta}^{n+1}$, $T^{n+1}$}
\SetKwFunction{size}{size}
\SetKwFunction{loadvectorized}{load\_vectorized}
\SetKwFunction{storevectorized}{store\_vectorized}
\SetKwFunction{timeintegration}{time\_integration}
$i \gets 0$\;
\While{$i <\ $ \size{$T^n$}}{
    \tcp{Load into vectorized array}
    $\begin{aligned}
    \check{X}_{\alpha_s}^n &\gets X_{\alpha_s}^n[i:i+n_\text{lanes}]\\
    \check{X}_{\alpha_m}^n &\gets X_{\alpha_m}^n[i:i+n_\text{lanes}]\\
    \check{X}_{\beta}^n &\gets X_{\beta}^n[i:i+n_\text{lanes}]\\
    \check{T}^n &\gets T^n[i:i+n_\text{lanes}]\\
    \check{T}^{n+1} &\gets T^{n+1}[i:i+n_\text{lanes}]
    \end{aligned}$
        
    \tcp{Solve local time integration problem}
    $\check{X}_{\alpha_s}^{n+1}$, $\check{X}_{\alpha_m}^{n+1}$, $\check{X}_{\beta}^{n+1}$ $\gets$ \timeintegration{$\check{X}_{\alpha_s}^{n}$, $\check{X}_{\alpha_m}^{n}$, $\check{X}_{\beta}^{n}$, $\check{T}^n, \check{T}^{n+1}$, $\Delta t$}\;

    \tcp{Store from vectorized array}
    $\begin{aligned}
     X_{\alpha_s}^{n+1}[i:i+n_\text{lanes}]&\gets \check{X}_{\alpha_s}^{n+1}\\
    X_{\alpha_m}^{n+1}[i:i+n_\text{lanes}] &\gets \check{X}_{\alpha_m}^{n+1}\\
    X_{\beta}^{n+1}[i:i+n_\text{lanes}] &\gets \check{X}_{\beta}^{n+1}\\
    {T}^{n+1}[i:i+n_\text{lanes}] &\gets \check{T}^{n}
    \end{aligned}$
    
    $i \gets i + v$\;
}
\end{algorithm}

\begin{algorithm}
\caption{Local explicit time integration of microstructure model on vectorized data.}\label{alg:explicit}
\DontPrintSemicolon
\newcommand{\cvm}{\check{\vec{m}}}

\KwData{$\cvm^n = [\check{X}_{\alpha_s}^n, \check{X}_{\alpha_m}^n]$, $\check{X}_{\beta}^n$, $\check{T}^n$, $\Delta t$}
\KwResult{$\check{X}_{\alpha_s}^{n+1}$, $\check{X}_{\alpha_m}^{n+1}$, $\check{X}_{\beta}^{n+1}$}
\SetKwFunction{size}{size}
\SetKwFunction{loadvectorized}{load\_vectorized}
\SetKwFunction{storevectorized}{store\_vectorized}
\SetKwFunction{computerates}{compute\_rates}
\SetKwFunction{instantaneouscorrections}{instantaneous\_corrections}

$\dot{\cvm}^n \gets$ \computerates{$\cvm^n$,$\check{T}^n$}\;

$\cvm^{n+1} \gets \cvm^{n} + \Delta t\,\dot{\cvm} $\;

$\cvm^{n+1}$, $\check{X}_{\beta}^{n+1}$ $\gets$ \instantaneouscorrections{$\cvm^{n+1}$, $\check{X}_{\beta}^n$, $\check{T}^{n}$}\;

\end{algorithm}

\begin{algorithm}
\caption{Local implicit time integration of microstructure model on vectorized data.}\label{alg:implicit}
\DontPrintSemicolon

\newcommand{\cvm}{\check{\vec{m}}}

\KwData{$\cvm^n = [\check{X}_{\alpha_s}^n, \check{X}_{\alpha_m}^n]$, $\check{X}_{\beta}^n$, $\check{T}^n$, $\check{T}^{n+1}$, $\Delta t$}
\KwResult{$\check{X}_{\alpha_s}^{n+1}$, $\check{X}_{\alpha_m}^{n+1}$, $\check{X}_{\beta}^{n+1}$}
\SetKwFunction{computerates}{compute\_rates}
\SetKwFunction{instantaneouscorrections}{instantaneous\_corrections}
\SetKwFunction{wrms}{weighted\_root\_mean\_square}

$\dot{\cvm}^n \gets$ \computerates{$\cvm^n$,$\check{T}^n$}\;
 
 $\cvm_0^{n+1}$, $\check{X}_{\beta,0}^{n+1}$ $\gets$ \instantaneouscorrections{$\cvm^{n}$, $\check{X}_{\beta}^n$, $\check{T}^{n+1}$}\;
 
 $\dot{\cvm}_0^{n+1} \gets$ \computerates{$\cvm_0^{n+1}$, $\check{T}^{n+1}$}\;
 
$i \gets 0$\;
\Repeat{$e < 1.0$}{
    $\cvm_{i+1}^{n+1} \gets \cvm^{n} + \frac{\Delta t}{2}(\dot{\cvm}_{i}^{n+1} + \dot{\cvm}^{n})$\;

    $\cvm_{i+1}^{n+1},\check{X}_{\beta,i+1}^{n+1} \gets  $\instantaneouscorrections{$\cvm_{i+1}^{n+1}$, $\check{X}_{\beta,i}^{n+1}$, $\check{T}^{n+1}$}\;

     $\dot{\cvm}_{i+1}^{n+1} \gets$ \computerates{$\cvm_{i+1}^{n+1}$, $\check{T}^{n+1}$}\;
    
    $e \gets$ \wrms{$\frac{\Delta t}{2}(\dot{\cvm}_{i+1}^{n+1} - \dot{\cvm}_{i}^{n+1})$}\;
    $i \gets i+1$\;
}

\end{algorithm}

Algorithm~\ref{alg:timeint} outlines the overall vectorized solution procedure for the microstructure model. Starting from the solution variables, $X_{\alpha_s}^n$, $X_{\alpha_m}^n$, $X_{\beta}^n$ and $T^n$, at the previous time $t_n$, the solution after a time increment $\Delta t$ with temperature $T^{n+1}$ is sought. We load a contiguous data slice from these five global vectors into vectorized arrays. Time integration is then  performed on the vectorized arrays. Afterward, the results at time $t_{n+1}$ are written back from the vectorized arrays into the respective global vectors. Note that the update of the temperature vector $T^{n+1} \gets T^n$ is also performed in this loop since the data is already loaded. Five load and four store operations must be performed for every evaluation point, totaling 72 bytes of memory transfer per evaluation point or 24 bytes per DoF (3 DoFs per evaluation point).

The explicit time stepping is outlined in Algorithm~\ref{alg:explicit}. The function \texttt{compute\_rates} evaluates the diffusion-based rates \eqref{eq:beta_to_alpha_s} -- \eqref{eq:alpha_s_to_beta}, and the function \texttt{instantaneuous\_corrections} evaluates instantaneous transformations between $\alpha_m$- and $\beta$-phase. Only a few arithmetic operations are needed for every set of vectorized arrays. In contrast, the implicit solution procedure shown in Algorithm~\ref{alg:implicit} usually requires fixed-point iteration and, thus, at least twice as many arithmetic operations as the explicit step. However, the amount of loaded and stored data is equal. As we will demonstrate in the examples, this typically leads to the explicit time integration scheme being memory-bound and the implicit scheme being compute-bound.

\subsection{Efficient approximation of transcendental functions}

\newcommand{\doubletype}{\texttt{double}}
\newcommand{\inttype}{\texttt{int64}}

The evolution equations of the microstructure model contain a few terms that necessitate the computation of a (non-integer) power or exponential function. Note that the power of a positive number can equivalently be written as
\begin{align}
    \label{eq:power_exp_log}
    a^x = \exp(x \ln a),\quad a>0,
\end{align}
which allows to compute the power of a number via a natural logarithm and an exponential function. 
At the time of writing, the C++ standard did not offer an implementation of these functions that could leverage SIMD hardware support. While copying and adapting the sophisticated implementations from the standard library or wrapping an existing library supporting vectorized data types would, in theory, be possible, we want to follow a different strategy here. The C++ standard implementation and most other libraries \cite{shibata2019sleef} must deal with a wide range of applications and consequently are implemented to be accurate up to machine precision. However, in the context of a numerical method, we accept a much less accurate result than machine precision. The chosen strategy embeds this fact by allowing the definition of an approximation that is accurate enough while minimizing the number of arithmetic operations.

The strategy employed in this work assumes the ubiquitous IEEE-754 standard \cite{ieee754} to represent floating point numbers. In this standard, a real number is represented as
\begin{align}
\label{eq:ieee754}
  (-1)^s 2^{p-b}(1+m).
\end{align}
For a 64-bit representation, i.e., the \doubletype{} data type in C/C++, the bias is set to $b = 1023$, the sign $s$ consumes a single bit, the exponent $p$ (an integer) consumes 11 bit, and the mantissa $m$ (a binary fraction) consumes 52 bit. The idea of directly computing and manipulating the bitwise representation was initially brought up in \cite{schraudolph1999fast}. Here, we follow the refined implementation discussed in \cite{malossi2015fast,perini2018fast}. Let us first find an approximation to the exponential function $z_{\exp}$ by rewriting,
\begin{align}
\label{eq:exp_integer_fraction}
    z_{\exp} := \exp(x) = 2^{x \log_2(e)} = 2^y =  2^{y_i}2^{y_f},
\end{align}
where $y_i = \lfloor x \log_2(e) \rfloor$ is an integer\footnote{The floor function $\lfloor x \rfloor$ returns the largest integer not exceeding $x$.}, and $y_f = x \log_2(e) - y_i \in [0,1)$ is a rational number.  By comparing \eqref{eq:exp_integer_fraction} to \eqref{eq:ieee754}, we find that
\begin{align}
    s &= 0,\\
    \label{eq:exp_integer_part}
    2^{y_i} &= 2^{p-b},\\
    \label{eq:exp_fractional_part}
    2^{y_f} &= 1+m.
\end{align}
The sign bit is always zero, as expected for exponentiation. The exponent $p$ can directly be computed from \eqref{eq:exp_integer_part} as 
\begin{align}
    \label{eq:exp_exponent}
    p = y_i + b.
\end{align}
%
%
By rearranging \eqref{eq:exp_fractional_part} and introducing a correction function $\mathcal{K}_{\exp}(y_f)$, we write:
\begin{align}
\label{eq:exp_correction_function_fractional}
     2^{y_f} &= 1+m =  1 + y_f -\mathcal{K}_{\exp}(y_f), \quad \text{with} \quad \mathcal{K}_{\exp}(y_f) = 1+y_f - 2^{y_f},
\end{align}
which leads to the mantissa
\begin{align}
\label{eq:exp_mantissa}
m = y_f - \mathcal{K}_{\exp}(y_f).    
\end{align}
The correction function $\mathcal{K}_{\exp}(y_f)$ is replaced with a polynomial approximation to circumvent the need to compute a (non-integer) power of 2. Note that this is the only approximation that is performed for the computation of the exponential function. The polynomial approximation can be tailored to the required accuracy by polynomial interpolation or regression. In this contribution, we use a least-squares fit of a polynomial of degree 7 to data sampled on 1000 equidistant points in $[0,1)$. The resulting coefficients are listed in Table \ref{tab:coefficients_approx_correction}. Tests revealed that adding more sample points does not increase the accuracy of the approximation in a relevant manner for our application. In contrast to \cite{perini2018fast}, we do not consider the derivatives of $\mathcal{K}_{\exp}(y_f)$ in the computation of the coefficients as we found their impact negligible.

\begin{table}
    \centering
    \caption{Polynomial coefficients for the approximated correction functions $\mathcal{K}_{\exp}(y_f)$  and $\mathcal{K}_{\ln}(y_f)$.}
    \label{tab:coefficients_approx_correction}
    \begin{tabular}{rS[table-alignment-mode=marker]S[table-format=1.12]}
    \toprule
          $i$ & \multicolumn{1}{l}{$a_i$ {in} $\mathcal{K}_{\exp}(y_f) = \sum_{i=0}^{n} a_iy_f^i$} & \multicolumn{1}{l}{$b_i$ in $\mathcal{K}_{\ln}(y_f) = \sum_{i=0}^{n} b_iy_f^i$}\\
          \midrule
         0&$\num{1.213071811889e-10}$&$\num{1.84775672096293e-10}$\\
         1&\num{3.068528102657e-1}&\num{1.44269504084132} \\
         2&\num{-2.40226342399359e-1}&\num{-0.721347520143005} \\
         3&\num{-5.55053313414954e-2}&\num{0.480898345526187} \\
         4&\num{-9.6135243288483e-3}&\num{-0.360675000332004} \\
         5&\num{-1.34288475963084e-3}&\num{0.288048466919235} \\
         6&\num{-1.43131744483589e-4}&\num{-0.235306287368882} \\ 
         7&\num{-2.1595656126349e-5}&\num{0.183102904829435} \\
         8 & &\num{-0.1209962689793}\\
         9 & &\num{0.0591503811592113} \\
         10 & &\num{-0.0181149492489989}\\
         11 & &\num{0.00254488675605743}\\
         \bottomrule
    \end{tabular}   
\end{table}
\begin{figure}
    \centering
    \includegraphics[width=\linewidth]{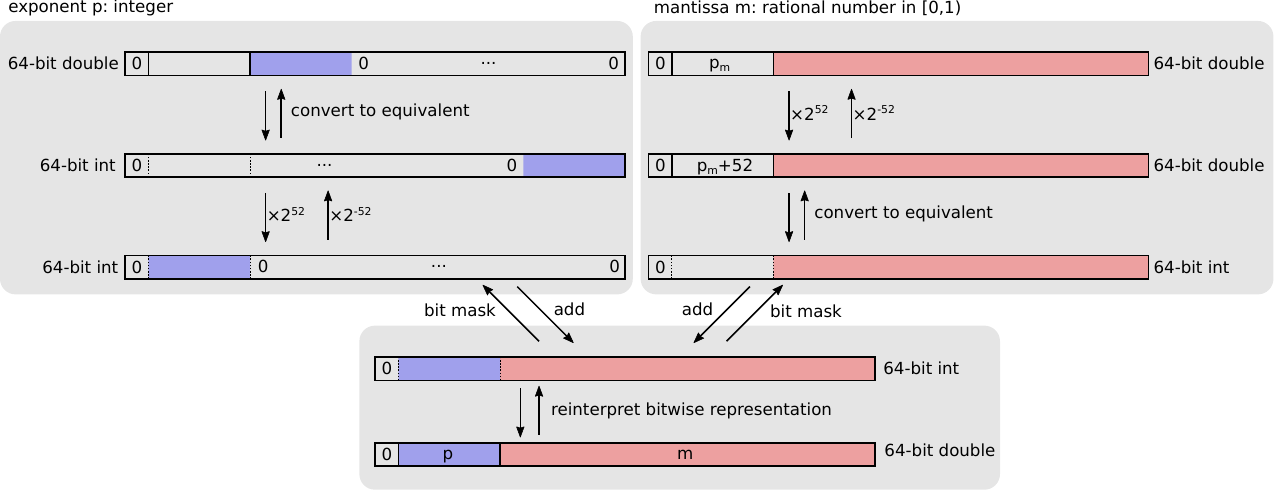}
    \caption{Following the arrows from top to bottom illustrates how to synthesize an IEEE754 \doubletype{} representation from a separate exponent $p$ (blue) and mantissa $m$ (red). Following the arrows from bottom to top shows the inverse operation, namely a decomposition of a \doubletype{} representation into exponent and mantissa. }
    \label{fig:bitwise_double}
\end{figure}
Since we determined the sign bit $s=0$, the exact exponent $p = y_i +b$, and the approximate mantissa $m = y_f -\mathcal{K}_{\exp}(y_f)$ of $z_{\exp}$ in \eqref{eq:exp_integer_fraction}, all that remains to be done is filling their bitwise representations into the IEEE754 conforming layout. An elegant way to achieve this can be derived by filling a 64-bit integer (\inttype{}) with the values of $s$, $p$, and $m$ and then interpreting the result as a (64-bit) \doubletype{} value. A graphical depiction of the approach is shown in Figure~\ref{fig:bitwise_double}. The exponent $p$ -- an integer stored inside a \doubletype{} representation -- is converted into the equivalent \inttype{} format and then multiplied with $2^{52}$, shifting it 52 bits to the left. Multiplying the mantissa $m < 1$ by $2^{52}$  shifts the decimal point behind the last (binary) digit, thus making it an integer, which is then converted into the equivalent \inttype{} format. In C++, the conversion from \doubletype{} to \inttype{} is achieved by a \texttt{static\_cast<int64>} operation.  The two \inttype{} numbers derived from $p$ and $m$ have no overlapping non-zero bits. They are added together to yield a combined \inttype{} with the bitwise representation of $z_{\exp}$ when reinterpreted as a \doubletype{} number( using \texttt{reinterpret\_cast<double>} in C++). Programmatically speaking, this algorithm can be written as:
\begin{align}
\label{eq:exp_bit_fill_1}
  z_{\exp} \leftarrow \texttt{reinterpret\_cast<double>}(
  2^{52}\times\texttt{static\_cast<int64>}(p)
  + \texttt{static\_cast<int64>}(2^{52}\times m) )
\end{align}
For an integer stored inside a \doubletype{} representation, we may flip the ordering of multiplication with another integer and a \texttt{static\_cast<int64>} to integer format. Thus, we can rewrite \eqref{eq:exp_bit_fill_1} as,
\begin{align}
\label{eq:exp_bit_fill_2}
  z_{\exp} \leftarrow \texttt{reinterpret\_cast<double>}(
  \texttt{static\_cast<int64>}(2^{52}\times (p + m) )),
\end{align}
and with \eqref{eq:exp_exponent} and \eqref{eq:exp_mantissa} after rearranging as
\begin{align}
\label{eq:exp_algorithm_1}
    y &\leftarrow x \log_2(e)\\
    \label{eq:exp_algorithm_2}
    y_f &\leftarrow y - \lfloor y \rfloor\\
    \label{eq:exp_algorithm_3}
    z_{\exp}
  &\leftarrow \texttt{reinterpret\_cast<double>}(
  \texttt{static\_cast<int64>}(A \times (y - \mathcal{K}_{\exp}(y_f)) + B ))
\end{align}
The coefficients $A = 2^{52}$, $B= 2^{52}\cdot1023$ and $\log_2(e)$ can be precomputed. The operations required in \eqref{eq:exp_algorithm_1}--\eqref{eq:exp_algorithm_3} are multiplication, addition, flooring, and type conversions. All of these are typically available in an instruction set for vector extensions. Thus, it is straightforward to implement \eqref{eq:exp_algorithm_1}--\eqref{eq:exp_algorithm_3} using SIMD in a given instruction set architecture.

The same ideas can be applied to derive a fast, vectorized approximation of the natural logarithm $\ln(x)$. We perform a change of basis and insert the IEEE754 double representation \eqref{eq:ieee754} to obtain:
\begin{align}
\label{eq:ln_approx}
    z_{\ln} = \ln(x) = \ln(2) \log_2(x) = \ln(2)\log_2\left[2^{p-b}(1+m)\right] = \ln(2)\lbrack \underbrace{(p-b)}_{l_i} + \underbrace{\log_{2}(1+m)}_{l_f} \rbrack.
\end{align}
Again, we identify an integer part $l_i$ and a fractional part $l_f$. The integer part $l_i$ is the exponent in the \doubletype{} representation of $x$. The fractional part $l_f$ is once more replaced by a correction function $\mathcal{K}_{\ln}(m) \approx \log_{2}(1+m) $ which takes the mantissa $m$ of $x$ as an argument. The exact form is replaced with an approximated polynomial of degree 10 determined via a least-squares fit of 1000 sample points in the relevant interval $m \in [0,1)$. The coefficients are given in Table \ref{tab:coefficients_approx_correction}. The exponent $p-b$ and the mantissa $m$ are extracted from $x$ by bit manipulations. Note that these operations are the inverse of the operations performed to synthesize a \doubletype{}, as shown in Figure~\ref{fig:bitwise_double} when following the arrows from the bottom to the top. Again, $\eqref{eq:ln_approx}$ is straightforward to implement with SIMD instructions.

To evaluate polynomials $\mathcal{K}_{\exp}$ of degree 7 and $\mathcal{K}_{\ln}$ of degree 10, we make use of Estrin's scheme \cite{estrin1960organization}. Although this scheme requires more floating point operations than the classical Horner scheme, it is better suited for the two separate fused multiply-add (FMA) units typically available on modern hardware since it allows independent computation of terms, thus shortening dependency chains. Applying this technique to the polynomial approximation of the correction function $\mathcal{K}_{\exp}$ yields,
\begin{align}
    \mathcal{K}_{\exp}(y_f) = 
        ((
          (a_7y_f + a_6)y_f^2 +
          (a_5y_f + a_4))y_f^4 +
          ((a_3 y_f + a_2)y_f^2+(a_1 y_f + a_0))),
\end{align}
where pairs of parentheses group expressions that can be computed by an FMA instruction.

\section{Numerical experiments}

The experiments are run on compute nodes consisting of two Intel Xeon Gold 6230 CPUs (total of 40 cores per compute node) running at $\num{2.1}\, \si{\giga\hertz}$ with a measured peak performance of $\num{2.5}\, \si{\tera Flops \per\second}$ and a DAXPY memory bandwidth of $\num{162}\ \si{\giga\byte\per\second}.$ The DAXPY benchmark updates a vector $\vec{y}$ in place according to $\vec{y} \leftarrow \vec{y} + \alpha\vec{x}$,  which, from a memory access perspective, is close to the microstructure model. A second setup consists of compute nodes with two AMD EPYC 9354 CPUs (total of 64 cores per compute node) running at 3.25\, \si{\giga\hertz} with a measured peak performance of $\num{3.75}\, \si{\tera Flops \per\second}$ and a DAXPY memory bandwidth of $\num{617}\ \si{\giga\byte\per\second}.$ All performance measurements are conducted and analyzed with the \texttt{LIKWID} suite \cite{treibig2010likwid}.

\subsection{Performance analysis}
\begin{figure}
    \centering
    \includegraphics[width=.49\linewidth]{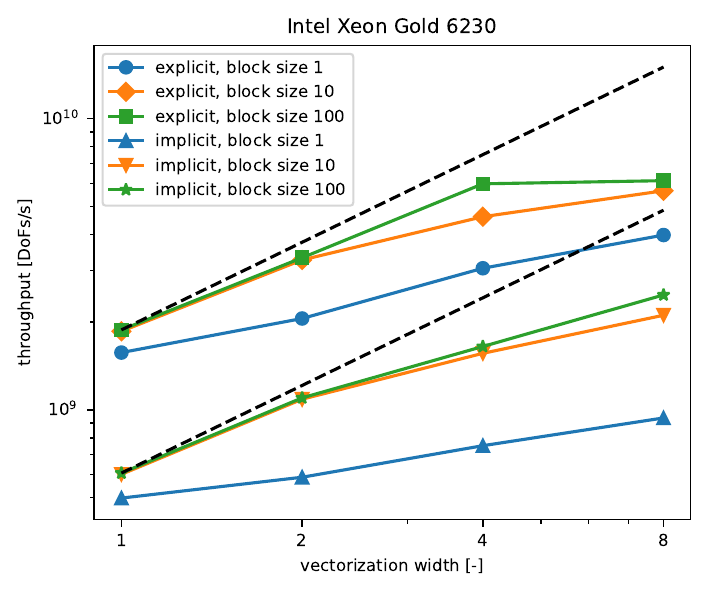}
    \includegraphics[width=.49\linewidth]{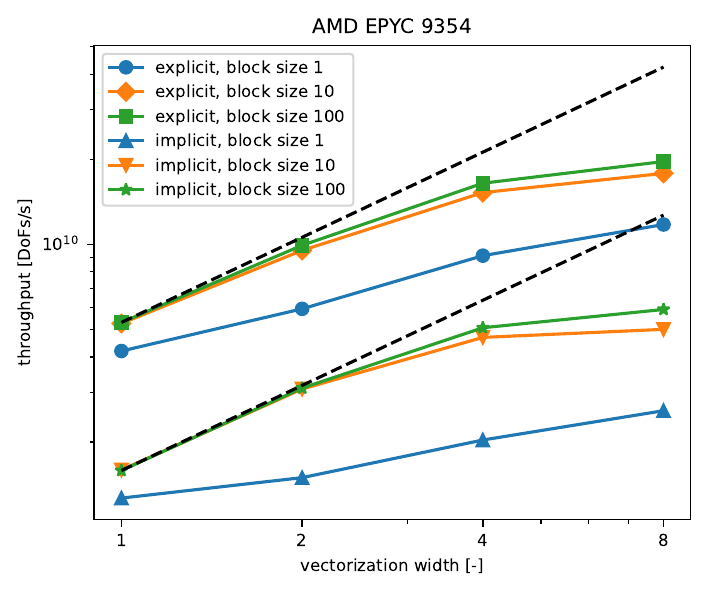}
    \caption{Impact of vectorization width on throughput of the microstructure model solved with an explicit or implicit scheme on Intel and AMD hardware.}
    \label{fig:solution_time_vectorization}
\end{figure}
\begin{figure}
    \centering
    \includegraphics[width=.49\linewidth]{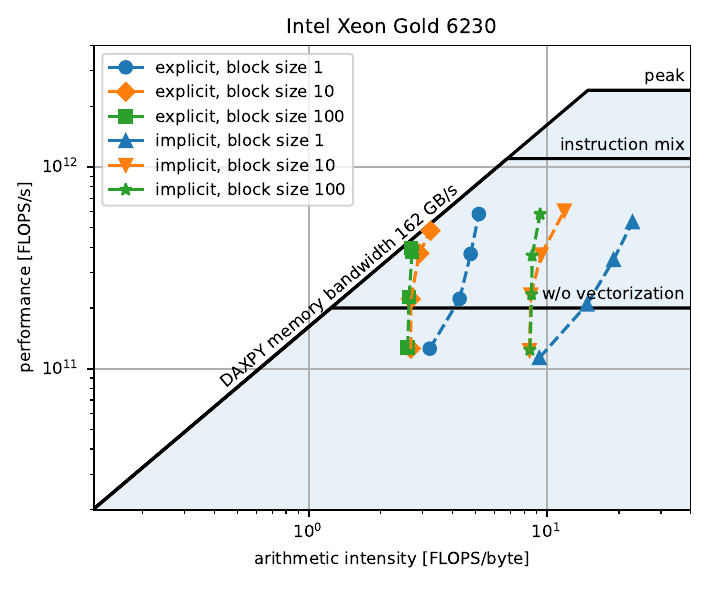}
    \includegraphics[width=.49\linewidth]{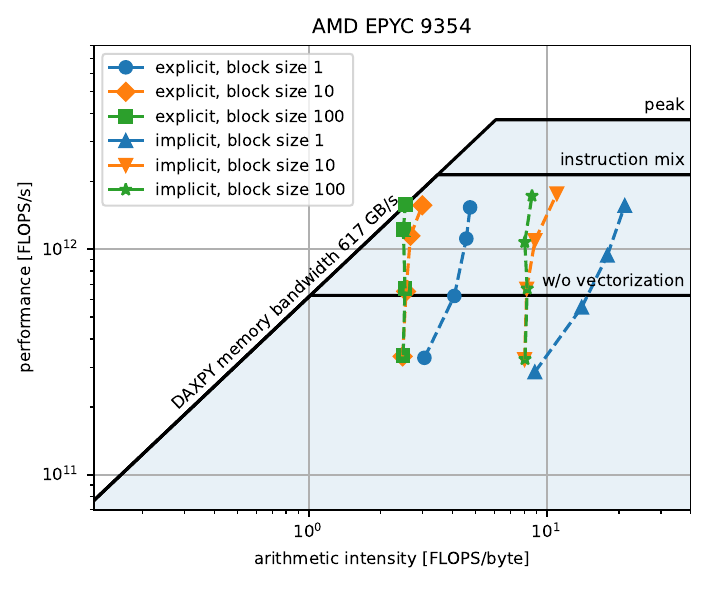}
    \caption{Roofline plot of a single time step performed in the microstructure model with an explicit or implicit scheme at different data block sizes. Data points move upwards along the dashed lines for increasing vectorization widths. }
    \label{fig:roofline_vectorization}
\end{figure}
As a first numerical study, we investigate the microstructural model implementation in isolation.
A significant challenge for a vectorized implementation of the microstructure equations lies in the conditional branches. As outlined above, we rely on the fact that spatially close material points (in the physical domain) likely need the same treatment. Such points are often located next to each other in a global vector of unknowns arising from spatial discretization. Three different synthetic sets of test data are used. The first test layout consists of alternating data, so neighboring entries in global vectors require different conditional branches in the microstructure model. The second consists of contiguous blocks of 10 entries, and the third consists of contiguous blocks of 100 entries. The throughput, defined as the number of DoFs divided by the solution time, is shown for these three block sizes and the explicit and implicit time integration scheme in Figure~\ref{fig:solution_time_vectorization}. Most importantly, the throughput increases for all block sizes when increasing the vectorization width. This holds for explicit and implicit time integration schemes. This behavior is achieved by implementing the expensive exponential and power functions to exploit SIMD instructions. Without this implementation, one would observe increased computation times and lower throughput for high vectorization widths at small block sizes. Due to unavoidable conditional computations in the case of block size 1, the performance improvement when increasing the vectorization width cannot scale perfectly here. For the Intel hardware, the flattening of the curve above \num{6} \si{\giga DoF\per\second} is in good agreement with the maximum theoretically possible throughput of $({162\, \si{\giga Byte\per\second}})/({24\, \si{Byte\per DoF}}) = \num{6.75}\,\si{\giga DoF\per\second}$ for a fully memory-bound code.

Figure~\ref{fig:roofline_vectorization} shows a roofline plot of the tested data layouts. For increasing vectorization width, data points move upwards and thus closer to the hardware limits. The explicit scheme quickly saturates the maximum memory bandwidth for larger block sizes. This explains why no speedup is visible when increasing the vectorization width from four to eight in Figure~\ref{fig:solution_time_vectorization}: the implementation is memory-bound, and higher vectorization widths cannot speed up the computation if memory transfer is the bottleneck. An improvement can only be expected if data transfer is minimized further or more work is performed for loaded data. One way to achieve this is subcycling, which performs multiple time steps on the same loaded data in place and only stores the result of the last subcycle. However, in a practical setting, the microstructure model already outperforms the thermal model by a significant factor, so further optimization in that direction is not investigated here. For block size 1, we observe a growing arithmetic intensity for increasing vectorization width. In this case, multiple branches must be computed in all lanes, although the result is only stored in some lanes. For larger block sizes, it becomes less likely to require different branch evaluations on a vectorized array, and thus, the number of (unproductive) arithmetic operations decreases.

%
%


Furthermore, we analyzed the generated machine code for the implicit scheme with the machine code analyzer of the LLVM project \cite{mca2024}. Analyzing the instruction mix reveals that the achievable performance is limited by dependency chains and expensive instructions that occupy the same execution ports as the productive floating point operations. In particular, executing the necessary instructions to compute diffusion from $\beta$ to $\alpha_s$ phase without any branching or load/store instructions reveals that only 43\% of the work performed on the two relevant execution ports of the Intel hardware contributes to the floating point operation metric. For the AMD hardware, the equivalent factor is 57\%. Therefore, we introduce an application-specific  roofline into the roofline plots, where the (unrealistic) peak performance is scaled down by the respective factor for the instruction mix. We achieve 58\% of the instruction mix performance on Intel Gold hardware and 82\% on AMD EPYC. 

\subsection{Cube geometry}
 \begin{table}
     \centering
     \caption{Processing parameters of cube example.}
          \label{tab:cube_parameters}
     \begin{tabular}{llll}
     \toprule
          Parameter & Description & Value & Unit \\
          \midrule
         $v_\text{scan}$ &  Laser scan speed & 960& $\si{\milli\meter\per\second}$ \\
         $d_h$ & Hatch spacing & 0.08 & \si{\milli\meter}\\
         $W_\text{eff}$ & Effective laser power & 180 & \si{\watt}\\
         $R$ & Laser beam radius & 0.06& \si{\milli\meter}\\
         $h_p$ & Powder layer thickness & 0.05  & \si{\milli\meter}\\
         \bottomrule
         \end{tabular}

 \end{table}

 \begin{figure}
     \centering
     \includegraphics[width=\linewidth]{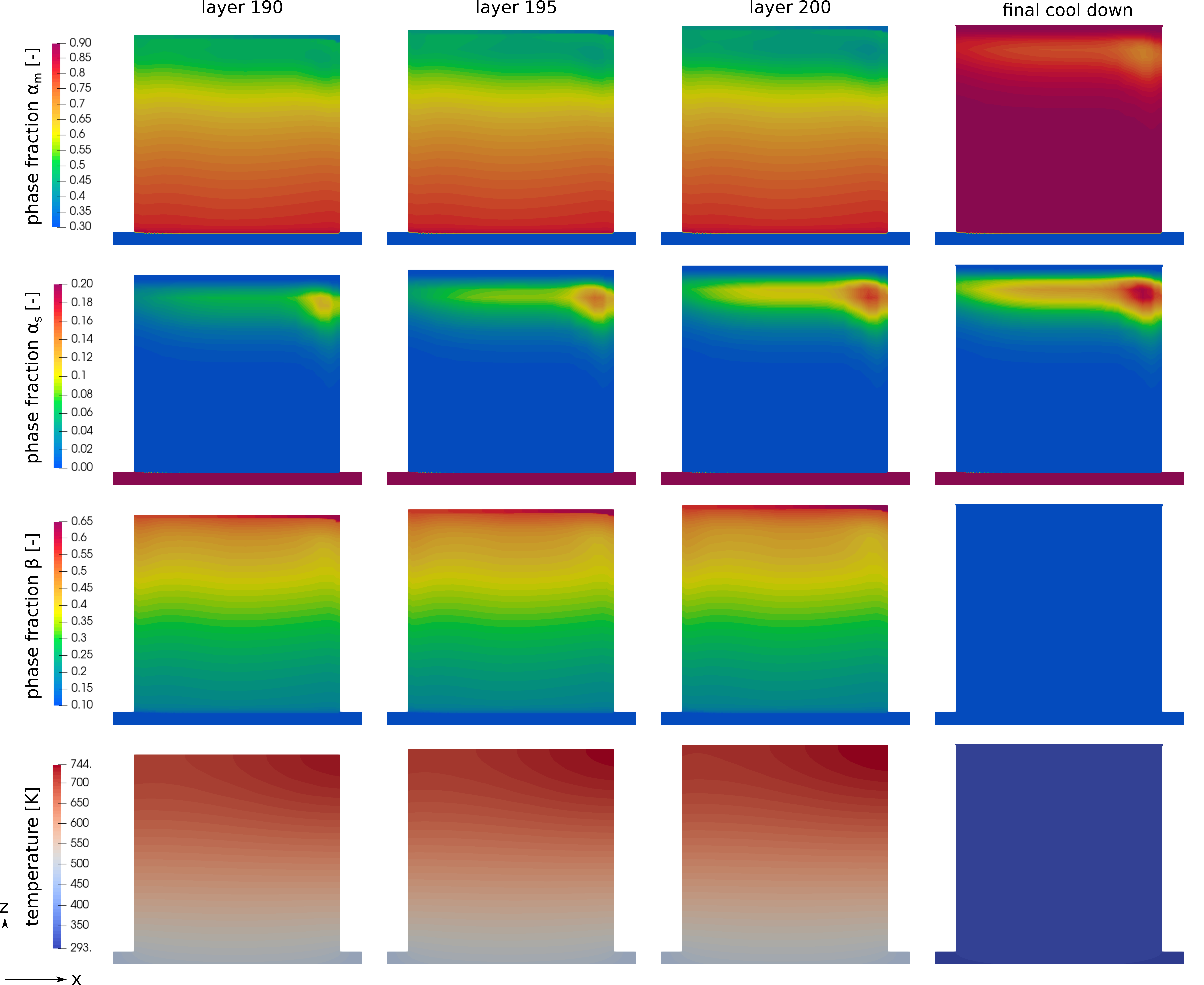}
     \caption{Phase fractions and residual temperature after processing and cooling of layers 190, 195, 200, and after a final cool down of the cube geometry. Results are depicted in a slice at $y=5\, \si{\milli\meter}$ and the base plate is cropped.}
     \label{fig:cube_phase_fractions_last_layers}
 \end{figure}

\begin{figure}
    \centering
    \includegraphics[width=\linewidth]{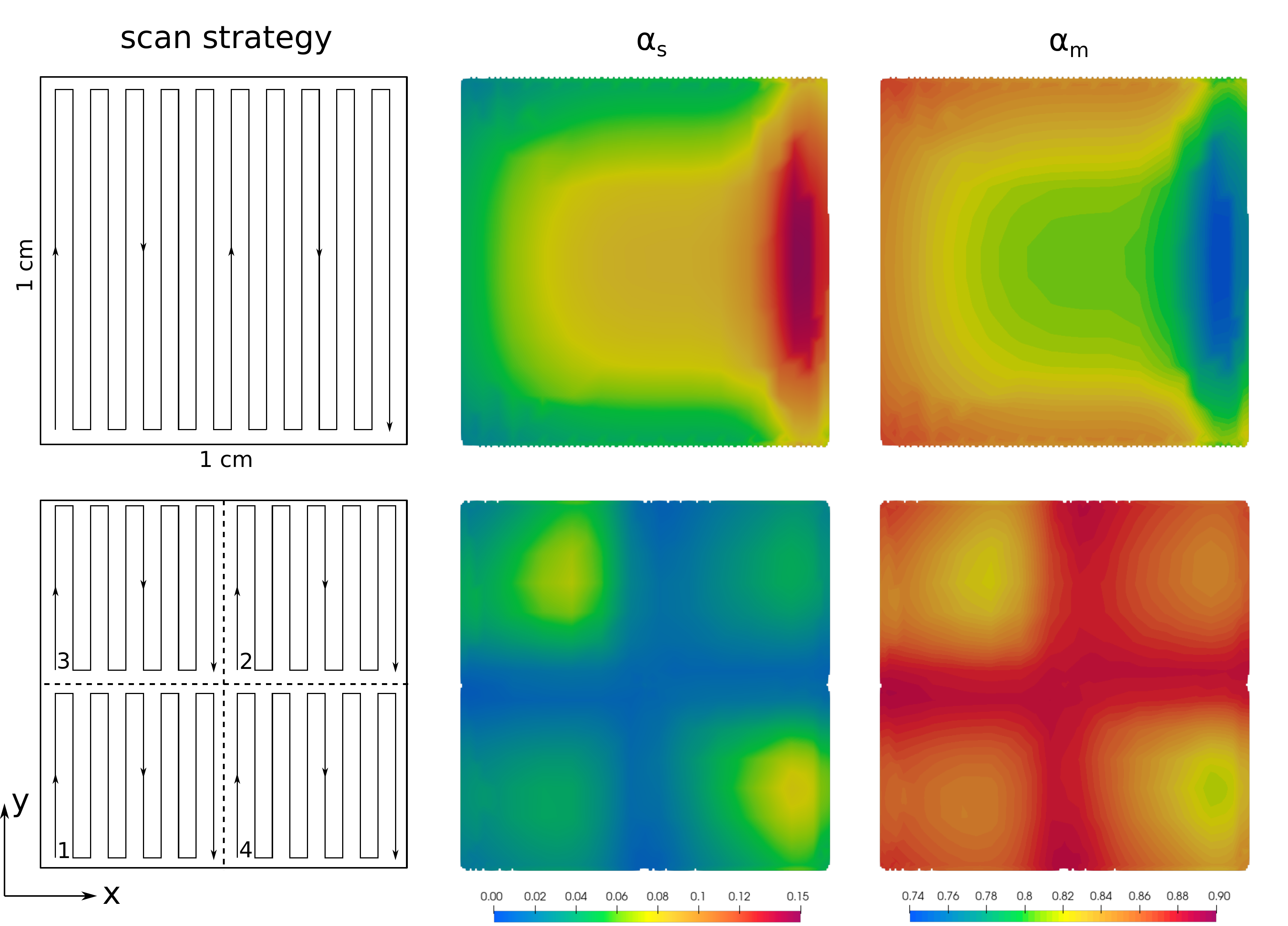}
    \caption{Final phase fraction of stable $\alpha_s$ and martensite $\alpha_m$ for cube example in a horizontal slice at $z=8\, \si{\centi\meter}$. Two different scan strategies are used to manufacture the cube: the first row shows the results for a continuous serpentine track extending over the cross-section. The second row shows the results for a scan path split into four islands, each containing a serpentine track.}
    \label{fig:cube_final_fractions_slice}
\end{figure}

\begin{figure}
    \centering
    \includegraphics[width=.5\linewidth]{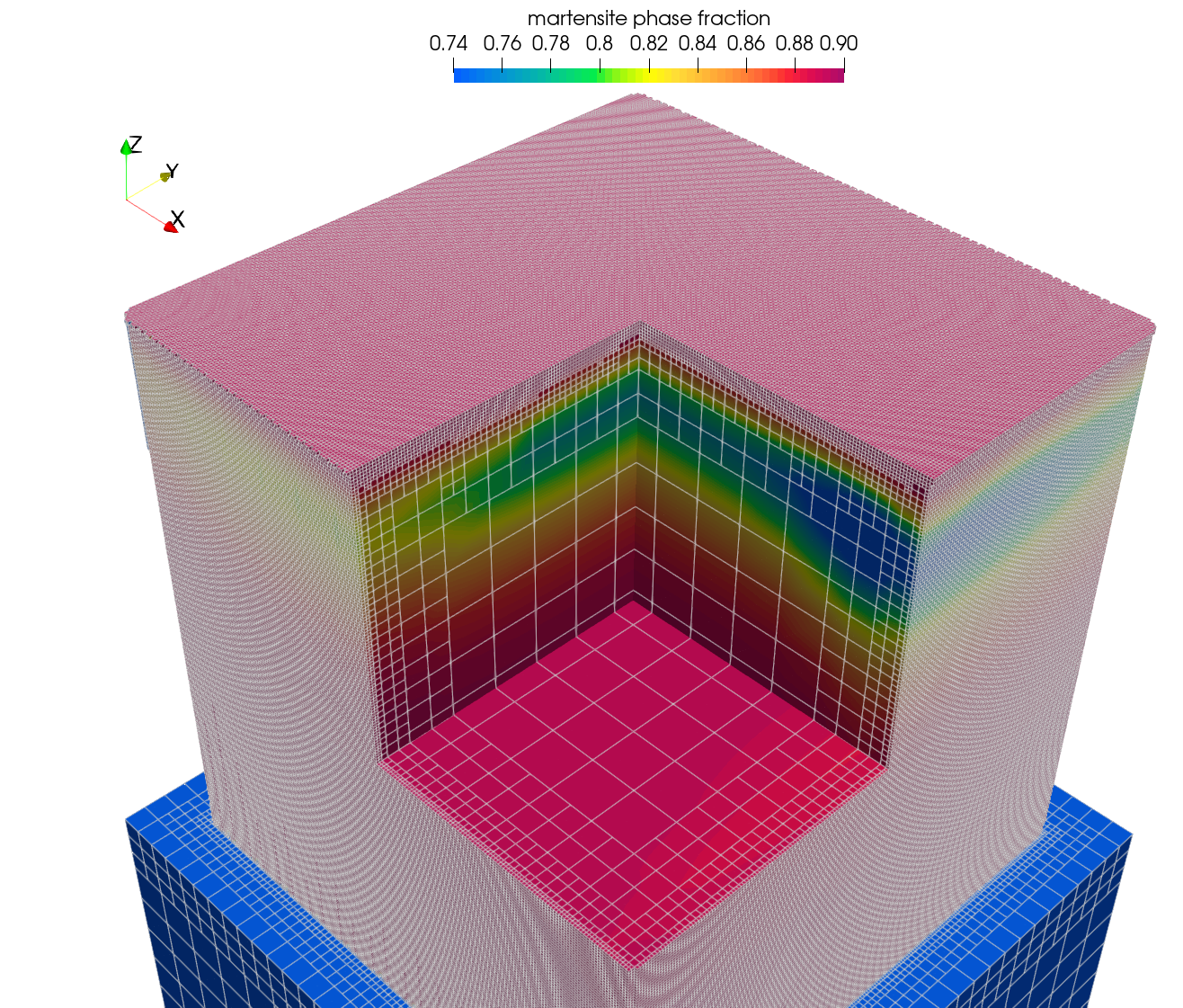}
    \caption{Final phase fraction of martensite $\alpha_m$ for cube example with single island scan track. An octant is cut out to show the asymmetric distribution of martensite induced by the asymmetric scan track. An adaptively refined mesh captures the built geometry.}
    \label{fig:cube_cutout_mesh}
\end{figure}

 In this example, we simulate manufacturing a $1 \times 1 \times 1\, \si{\cubic\centi\meter}$ cube from 200 powder layers with thickness $\num{0.05}\, \si{\micro\meter}$. The cube is placed on a base plate of dimensions $1.2 \times 1.2 \times 1\, \si{\cubic\centi\meter}$. The processing parameters are summarized in Table~\ref{tab:cube_parameters}. A time step size of $\Delta t_\text{active} = \num{2e-5}\,\si{\second}$ is used in the active laser phase and after every layer, a cool down phase of $\num{1}\, \si{\second}$ is simulated. The first 2,000 steps of this phase are simulated with the same time step size as the active phase. Afterward, the macroscopic time step size is doubled after every ten time steps for the thermal model up to a maximum step size of $0.1\, \si{\second}$. The microstructure model uses the identical time step size as the thermal model if the step size is less than or equal to $\num{1e-2}\, \si{\second}$ and, otherwise, uses the subcycling technique described earlier. After processing the last layer, the geometry is cooled to room temperature over $\num{100}\,\si{\second}$.

 As a first example, every layer is processed as a single island consisting of a continuous serpentine track, where the laser beam moves in $y$-direction and hatching proceeds in $x$-direction. The three phases and the residual temperature after layers 190, 195, and 200 are processed are shown in Figure~\ref{fig:cube_phase_fractions_last_layers}. A noticeable asymmetry can be seen in the higher layers, where the accumulated heat produces a.  Due to the long and continuous serpentine track, heat accumulates as the track hatches progress in positive $x$-direction, leading to lower cooling rates and decreased martensite formation and, instead more stable $\alpha_s$ phase formation in layers 170 to 190. This happens because the residual temperature only falls slightly below the martensite start temperature in higher layers. The material is held at elevated temperatures, giving enough time for the diffusion-driven formation of stable $\alpha_s$-phase. Due to the time required for a significant formation of $\alpha_s$-phase, the formation happens a few layers below the currently processed layer.  Note that the final cool down stage is essential to obtain the actual phase composition close to room temperature. During this stage, the already formed stable $\alpha_s$-phase remains. In the highest layers, the cooling rate is now large enough because no heat is added above layer 200. Consequently,  most of the  $\beta$-phase remaining after processing of layer 200 transforms into $\alpha_m$-phase b.  At room temperature, the $\beta$-phase fraction reaches its equilibrium value of 10\%.
 
 The cube test geometry is well-suited to study the effects of different scan strategies on the resulting microstructural composition. In addition to the single island scan track consisting of a continuous serpentine track, we investigate a scan track split into four disjoint islands, each consisting of a serpentine track. The tracks are shown in Figure~\ref{fig:cube_final_fractions_slice} along with the  stable $\alpha_s$- and martensite $\alpha_m$-phase fractions after the final cool down in  a horizontal slice at $z=8\, \si{\centi\meter}$. A strong asymmetry is visible in the in-plane distribution of the $\alpha_s$- and $\alpha_m$-phase fractions for the single-island scan strategy.
On the other hand, the four-island scan strategy produces a more homogeneous distribution with a higher average martensite fraction than the single-island scan. The different phase distribution is caused exclusively by the scan strategy, as all other parameters are identical. This observation again shows the need for scan-resolved models, such as the one presented in this work.

For a better idea of the different scales involved in a resolved part-scale simulation, a view of the adaptive mesh is shown in Figure~\ref{fig:cube_cutout_mesh}. The smallest cells' heights correspond to one powder layer thickness. To capture the geometry, the mesh is more refined close to the cube's surface. One could employ a boundary-fitted mesh for this simple geometry to save DoFs, as done in our previous work \cite{proell2023highly}.

 \begin{figure}
     \centering
     \includegraphics[width=\linewidth]{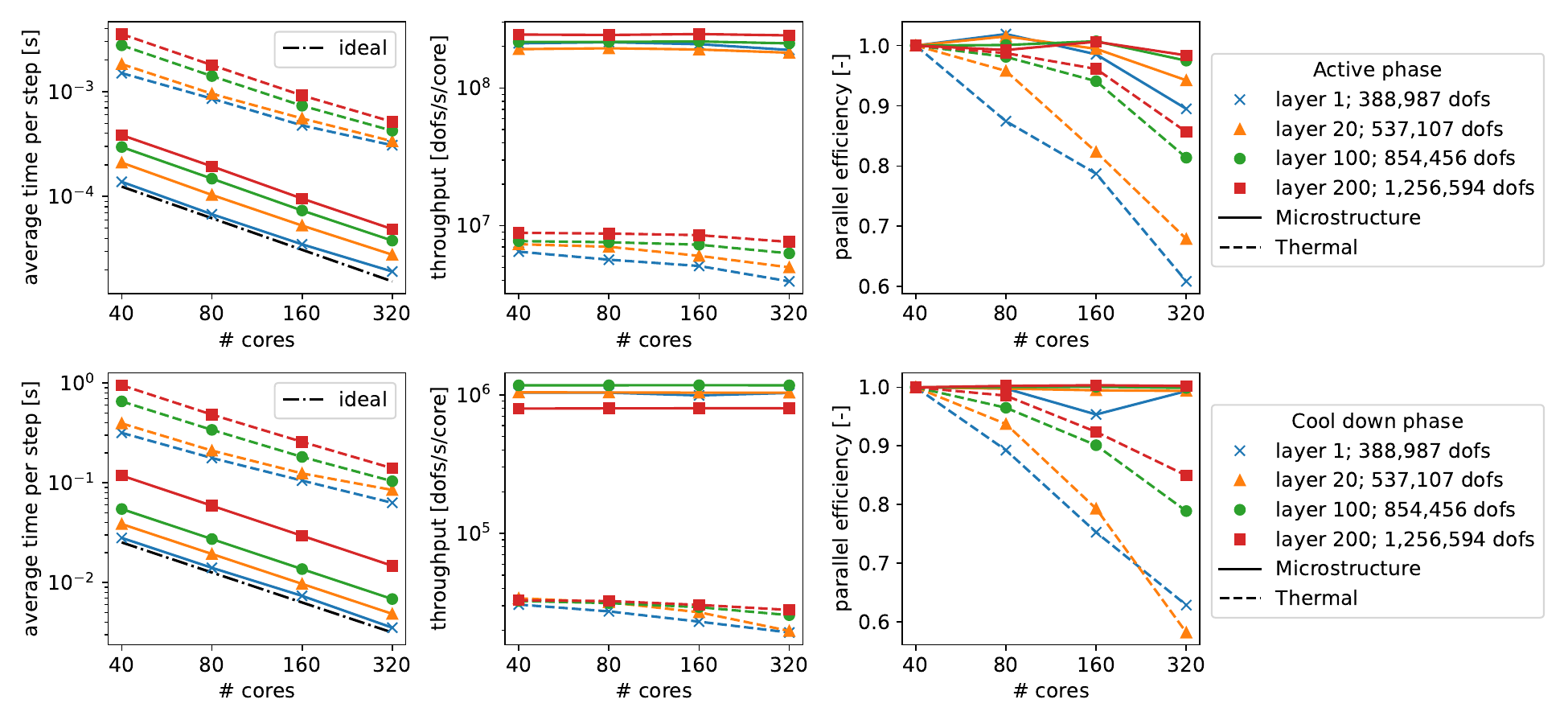}
     \caption{Strong scaling study of cuboid example with adaptive mesh coarsening. The first row shows the average solution time per step, the throughput, and the parallel efficiency in the active laser phase; the second row shows the same metrics in the cool down phase. Note that the reported time per step includes subcycling for the microstructure model in the later cool down steps. The number of DoFs is given for the thermal problem; the microstructure problem carries three times this number.}
     \label{fig:cuboid_strong_scaling_coarse}
 \end{figure}

 Finally, we also want to judge the performance of our implementation in this more practical example. The scenario with the single-island scan strategy is simulated in a strong scaling study on 1, 2, 4, and 8 nodes of the Intel Gold hardware mentioned in the introduction to this section connected by an Infiniband FDR (56 Gbit/s) interconnect. The resulting average time per time step, throughput, and parallel efficiency are shown in Figure \ref{fig:cuboid_strong_scaling_coarse} for the thermal and microstructure models separately. Here, parallel efficiency is defined as:
 \begin{align}
     \text{parallel efficiency} &= \frac{\text{reference compute time} \times \text{reference number of cores}}{\text{scaled compute time} \times \text{scaled number of cores}}
 \end{align}
Overall, the explicit and implicit solution of the microstructure equations requires around 10\% of the time of the respective thermal model. Note that the microstructure problem carries three times the number of DoFs as the thermal problem; thus, the throughput is further increased compared to the thermal problem. The microstructure implementation exhibits excellent parallel scalability, even in the first layer. This result is to be expected since no communication between processes is involved.
The total wall time when running the example on eight Intel Gold nodes is $\num{2.32}\,\si{\hour}$.

\subsection{Cantilever}
 \begin{table}
     \centering
     \caption{Processing parameters of cantilever example.}
          \label{tab:cantilever_parameters}
     \begin{tabular}{llll}
     \toprule
          Parameter & Description & Value & Unit \\
          \midrule
         $v_\text{scan}$ &  Laser scan speed & 960& $\si{\milli\meter\per\second}$ \\
         $d_h$ & (Approximate) hatch spacing & 0.11 & \si{\milli\meter}\\
         $W_\text{eff}$ & Effective laser power & 180 & \si{\watt}\\ 
         $R$ & Laser beam radius & 0.06& \si{\milli\meter}\\
         $h_p$ & Powder layer thickness & 0.04  & \si{\milli\meter}\\
         \bottomrule
         \end{tabular}

 \end{table}

\begin{figure}
     \centering
     \includegraphics[width=.7\linewidth]{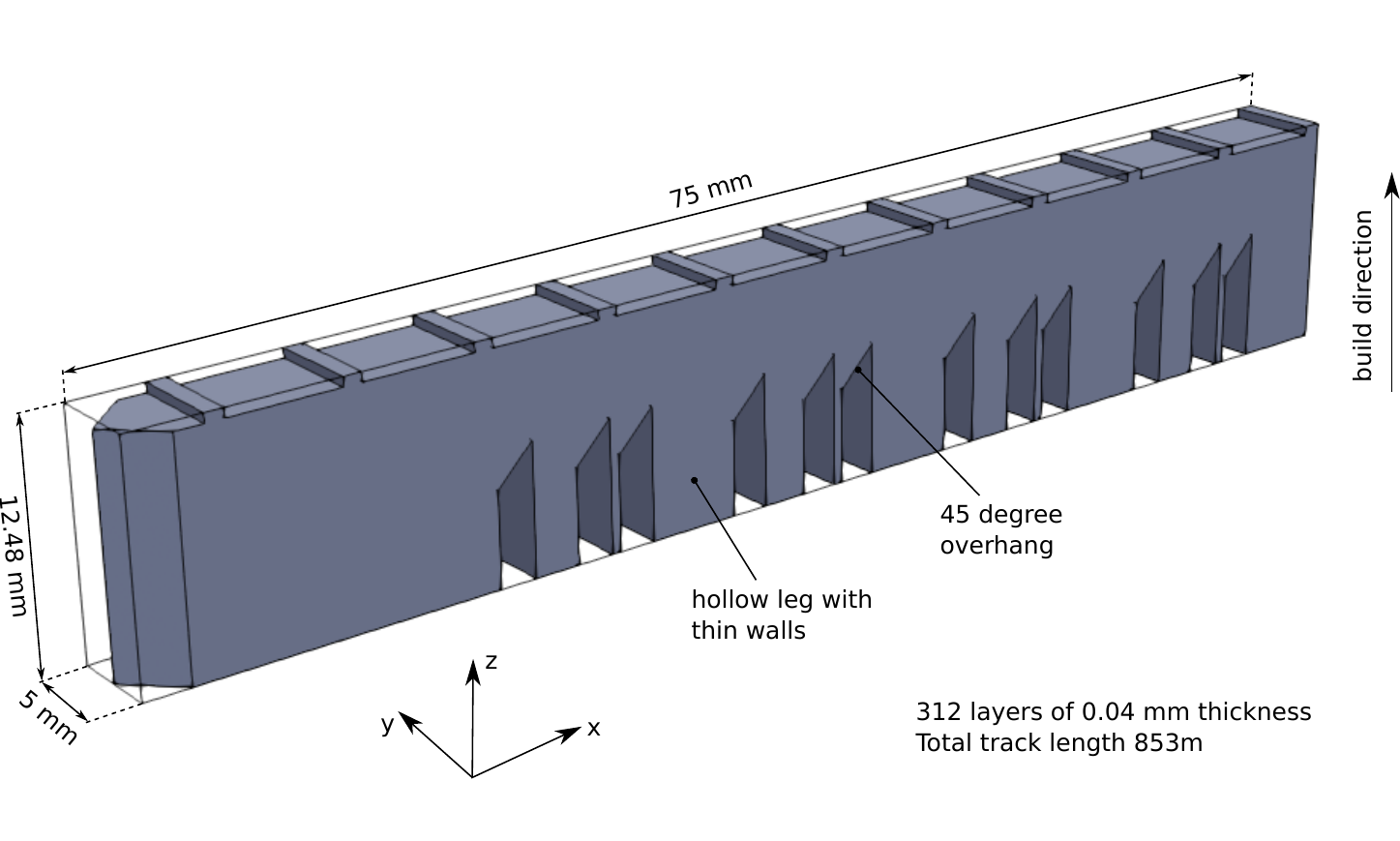}
     \caption{NIST AMBench 2022 cantilever geometry.}
     \label{fig:cantilever_overview}
 \end{figure}

As a last example, we investigate the well-known NIST AMBench cantilever geometry \cite{AMbench2022}. The geometry, shown in Figure~\ref{fig:cantilever_overview}, features thin-walled legs and overhang regions, leading to different local cooling rates. The geometry is built on a 10.56 \si{\milli\meter} high base plate section. To investigate the effect of different pre-heating temperatures, the bottom of the base plate is constrained to a fixed temperature $\hat{T} \in \lbrace 293\, \si{\kelvin}, 500\, \si{\kelvin},550\, \si{\kelvin}, 600\, \si{\kelvin}\rbrace$. The scan strategy is directly taken from \cite{AMbench2022}.	
The active scan phase in every layer is followed by a cool down phase of $1 \si{\second}$, which uses identical time step sizes as described for the cube example. After simulating all 312 layers, a final cool down of $\num{100}\,\si{\second}$ is simulated, during which the temperature on the bottom of the baseplate is set to $293\,\si{\kelvin}$ room temperature. The scan parameters are given in Table~\ref{tab:cantilever_parameters}.
 \begin{figure}
     \centering
     \includegraphics[width=\linewidth]{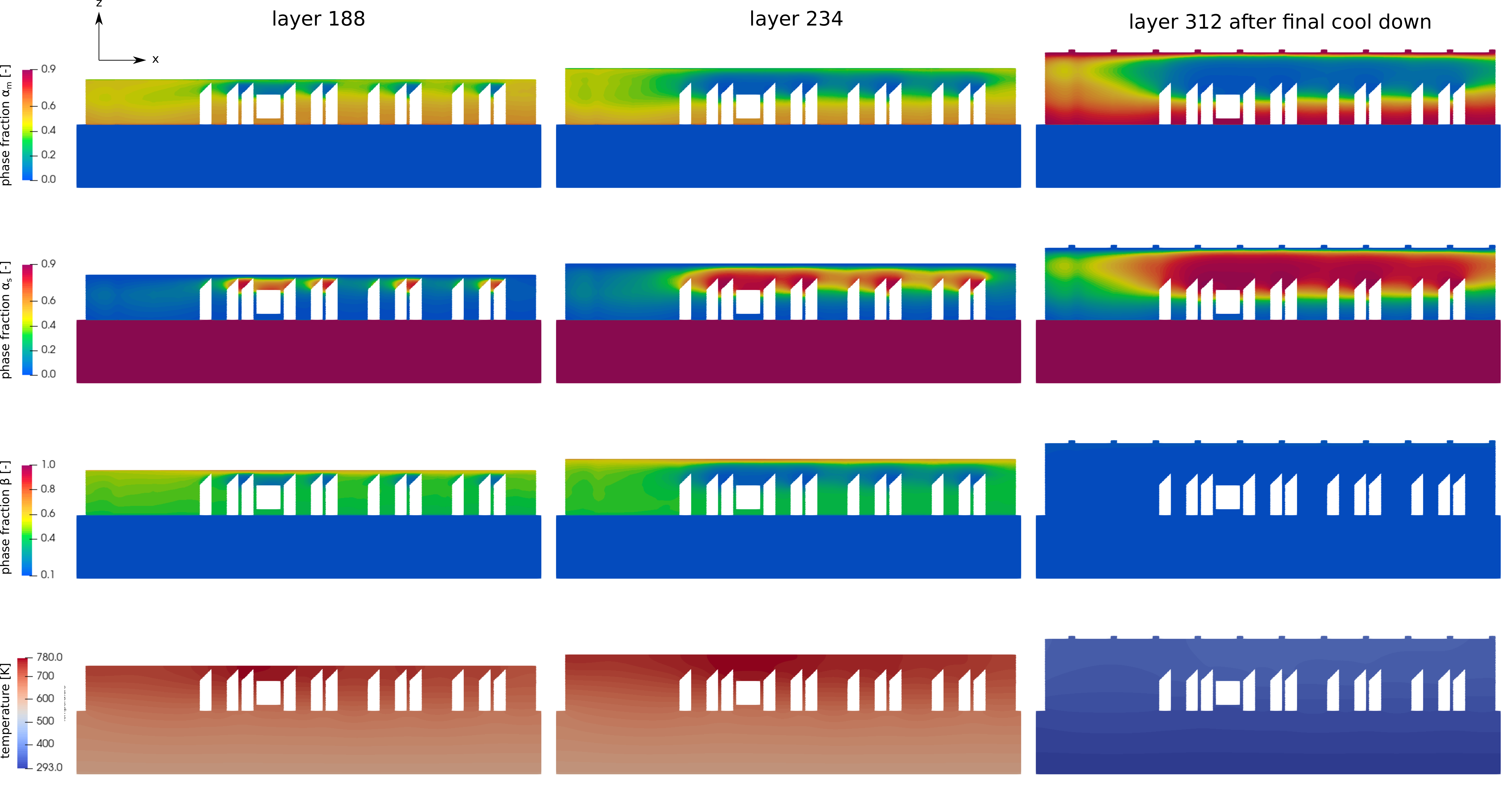}
     \caption{Phase fractions and residual temperature after processing and cooling of layers 188, 234, and 312 (including final cool down time). Results are depicted in the symmetry $xz$-plane of the cantilever geometry.}
     \label{fig:cantilever_phase_fractions_600K}
 \end{figure}
 \begin{figure}
     \centering
     \includegraphics[width=.7\linewidth]{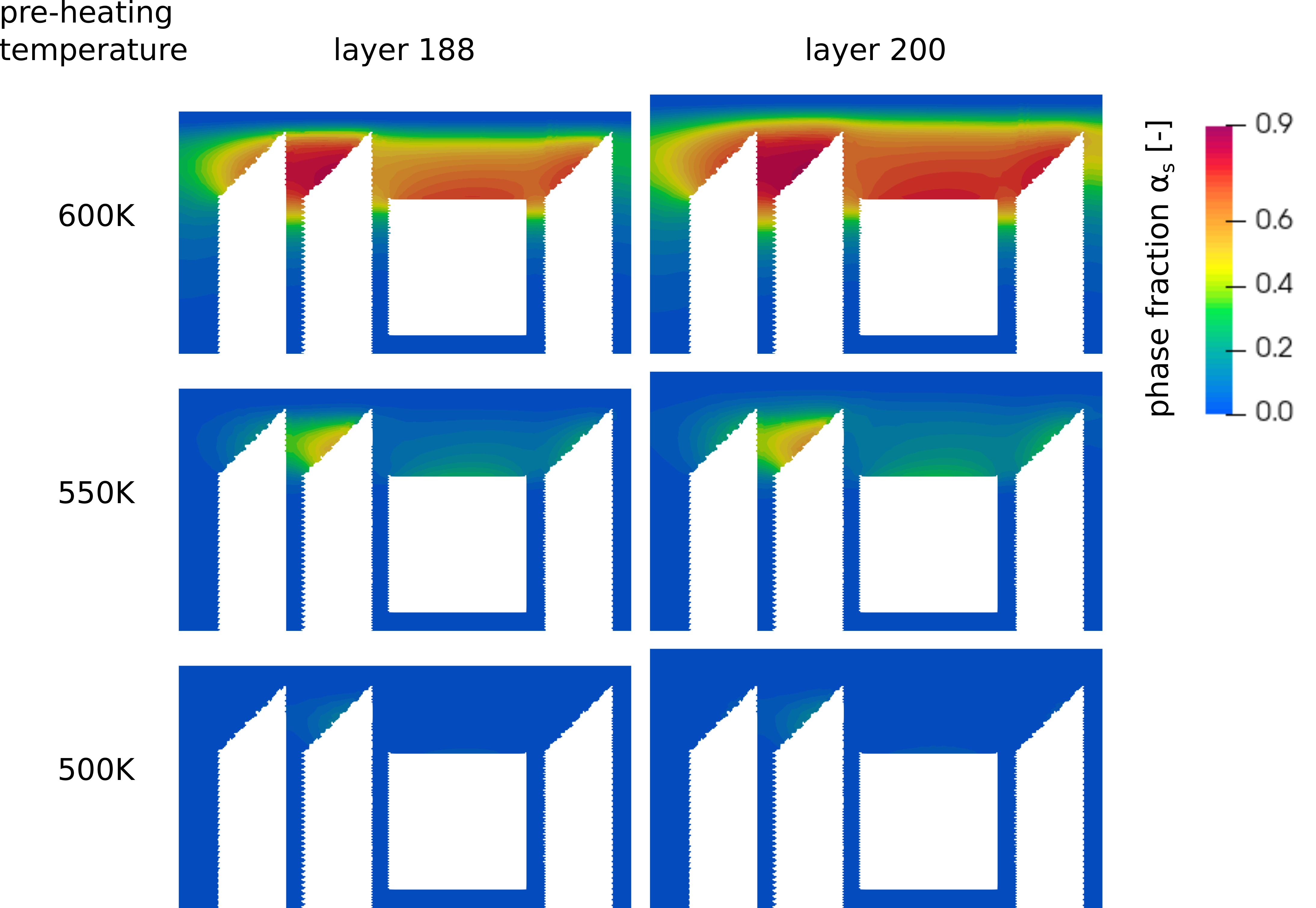}
     \caption{Detailed view of $\alpha_s$-phase fractions for different preheating temperatures after processing and cooling of layers 188 and 200. Results are depicted in the symmetry $xz$-plane of the cantilever geometry near the hollow leg structure.}
     \label{fig:cantilever_compare_preheating}
 \end{figure}
 \begin{figure}
     \centering
     \includegraphics[width=\linewidth]{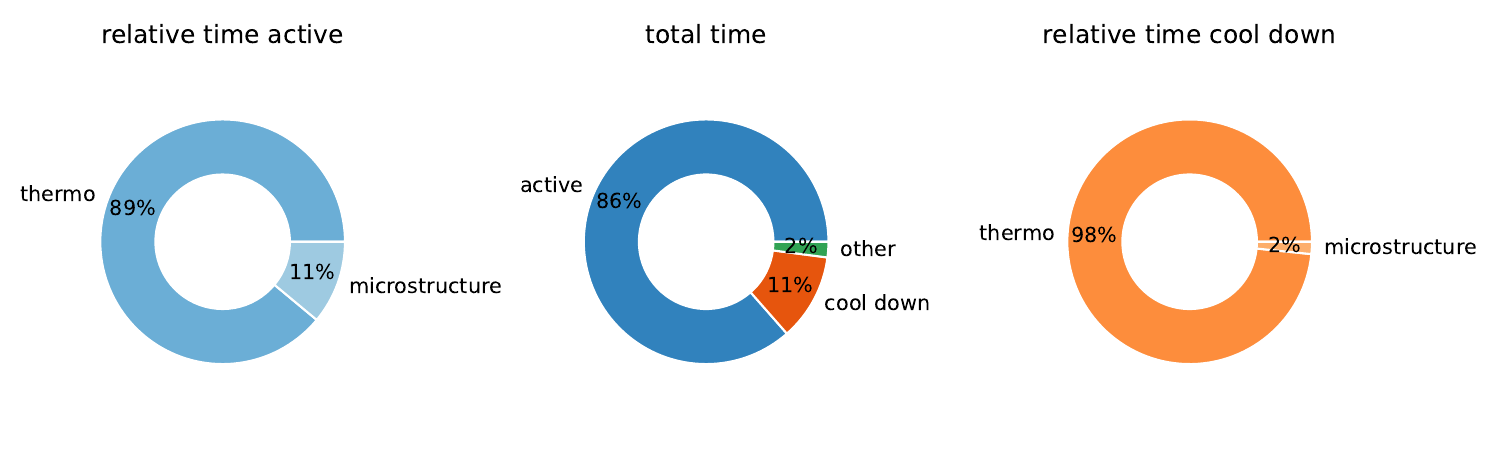}
     \caption{Distribution of total solution time over different parts of the solution procedure.}
     \label{fig:cantilever_epyc_time_split}
 \end{figure}

The case of $\hat{T} = 293 \si{\kelvin}$ leads to a fully martensitic microstructure and is not shown in more detail. Instead, we focus on the results for higher pre-heating temperatures. Figure \ref{fig:cantilever_phase_fractions_600K} shows the phase fractions and residual temperatures at various points in time for a pre-heating temperature $\hat{T} = 600\, \si{\kelvin}$. In the overhang regions and above the legs, a substantial amount of $\alpha_s$-phase forms over time due to the reduced cooling provided by the thin-walled legs. Again, the phase fraction $\alpha_s$ increases visibly a few layers after initial processing. It keeps growing until reaching 90\% stable $\alpha_s$-phase, the equilibrium value. The highest layers are not held at an elevated temperature for a sufficient time, so almost no $\alpha_s$-phase forms here and the microstructure is fully martensitic after the final cooling step.

While we varied the scan pattern in the last example, we now vary the pre-heating temperature and show the evolution of $\alpha_s$-phase for $\hat{T} \in \lbrace 500\,\si{\kelvin}, 550\,\si{\kelvin}, 600\,\si{\kelvin}\rbrace$ in Figure \ref{fig:cantilever_compare_preheating}. The final microstructure composition is sensitive to the pre-heating temperature, the scan track, and the interlayer cool down time.
 
The full simulation of the case with $\hat{T} = 293\, \si{\kelvin}$ takes 52.3\, \si{\hour} on four AMD Epyc nodes. Figure~\ref{fig:cantilever_epyc_time_split} breaks down the total solution time into active and cool down phase and the thermal and microstructure problem. Due to the high degree of optimization of the microstructure implementation, the coupled thermo-microstructure problem is obtained at only marginally increased computation time compared to the thermal problem alone.

\section{Conclusion}

We presented a highly efficient implementation of a coupled thermal-microstructure model. The proposed approach enables simulations on the scale of real parts with a scan-track resolved heat source. As demonstrated in the examples, considering the actual scan track is vital to capture variations in the microstructural composition caused by the scan strategy rather than the geometry. Simulations with hundreds of layers are possible in a few hours to a few days, depending on the build volume. While the investigated geometries are of a relevant scale, the presented methodology may be combined with layer-wise heating approaches in areas where this simplification is applicable to tackle scales beyond decimeters in future research.

The microstructure model equations contain conditional branches and computationally expensive mathematical functions. Through special approximations and a careful data layout, the proposed methodology can utilize modern hardware capabilities efficiently. The evaluation of the thermo-microstructure model comes with less than a 10\% increase in run time compared to the thermal model.

In future investigations, the current model may be refined to include homogenized information on the anisotropic orientation of grains induced by the thermal gradients. The model can then serve as the basis for microstructure-informed solid mechanics simulations. 

\section*{Acknowledgements}

The authors thank Maximilian Bergbauer for valuable discussions about performance modeling. Furthermore, the authors thank Neil Hodge, Jonas Nitzler, and Nils Much for their initial work on the microstructure model.

\appendix

\section{Analytical solution for initiating diffusion-based transformations}
\label{appendix:analytical_solution}

When initiating the diffusion process $\alpha_s \rightarrow \beta$ from an initial state $X_{\alpha_s} = 0.9$, $X_\beta = 0.1$, the rate form \eqref{eq:alpha_s_to_beta} cannot be used in combination with an explicit scheme since it always yields zero for the initial values. Instead, an approximate solution is used for the initial time steps when diffusion starts.

The Crank-Nicolson integration scheme with a fixed point iteration would not face this issue if one were to perturb the initial guess. However, for a unified implementation, the analytical solution is also used to initiate diffusion when using the Crank-Nicolson scheme.

\begin{figure}
   \centering
   \includegraphics[width=.75\linewidth]{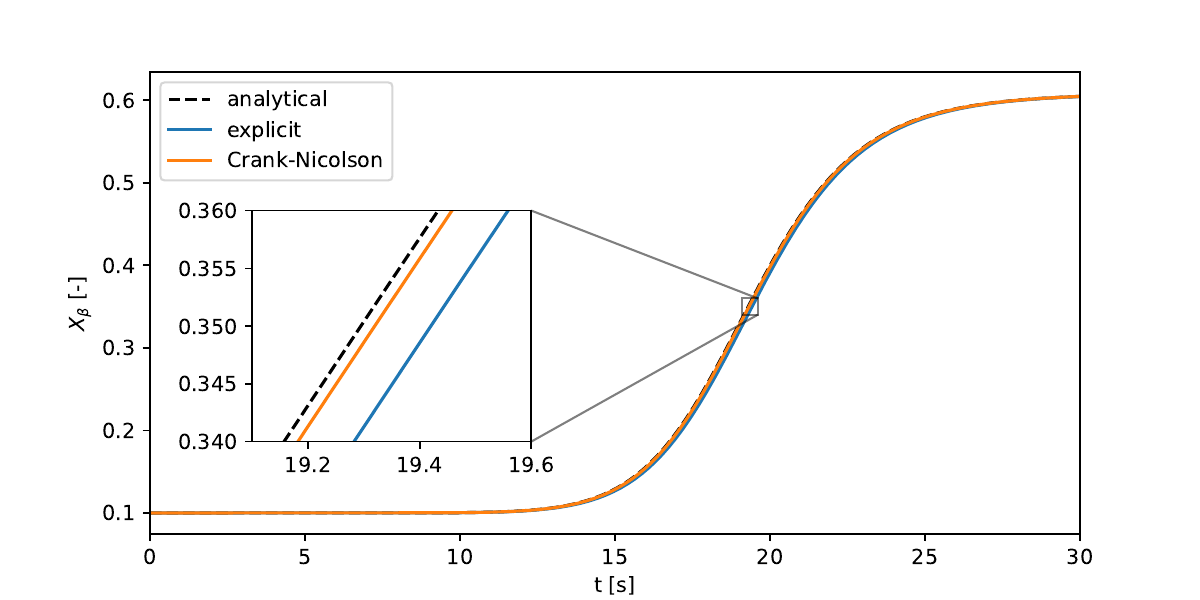}
   \caption{Comparison of analytical solution for transformation $\beta \rightarrow \alpha_s$ to explicit and Crank-Nicolson time integration.  The solution is computed for a constant temperature $T=1200\, \si{\kelvin}$ and initial values $X_{\alpha_s} = 0.9$ and $X_\beta = 0.1$.}
   \label{fig:analytical_solution_xb_verification}
\end{figure}

To derive the approximate analytical solution, we rewrite \eqref{eq:alpha_s_to_beta}, with the help of \eqref{eq:phase_continuity_alpha_beta} and the relations $X_\beta^\text{eq} = 1 -X_\alpha^\text{eq}$ and $\tilde{X}_\beta := X_\beta -0.1$ as
\begin{align}
    \dot{X}_\beta &= k_\beta(T) (0.9 - X_\alpha)^{\frac{c_\beta - 1}{c_\beta}}(X_\alpha - X_{\alpha}^\text{eq})^{\frac{c_\beta+1}{c_\beta}}\nonumber\\
    &= k_\beta(T) (\tilde{X}_\beta)^{\frac{c_\beta - 1}{c_\beta}}(X_\beta^\text{eq} - 0.1 - \tilde{X}_{\beta})^{\frac{c_\beta+1}{c_\beta}}.
\end{align}
We perform the substitution $\xi := \tilde{X}_\beta/(X_\beta^\text{eq} - 0.1)$ and obtain
\begin{align}
    \dot{\xi} = \underbrace{(X_\beta^\text{eq} - 0.1)k_\beta(T)}_{\tilde{k}} (\xi)^{\frac{c_\beta-1}{c_\beta}}(1-\xi)^{\frac{c_\beta+1}{c_\beta}},
\end{align}
which, assuming that $\tilde{k} = \text{const.}$, has a known solution given in \cite{avramov2014generalized}. For an initial value $\xi_n$ at $t_n$, the solution in the next step $t_n+\Delta t$ can be found as:
\begin{align}
\label{eq:analytical_solution_beta}
    \tilde{X}_{\beta}^{n+1} = (X_\beta^\text{eq} - 0.1) \left[1 + \left(\frac{c_\beta}{(X_\beta^\text{eq} - 0.1)k_\beta(T)\Delta t + c_\beta\left(\frac{\xi_n}{1-\xi_n}\right)^{\frac{1}{c_\beta}}} \right)^{c_\beta}\right]^{-1}
\end{align}
For numerical reasons, this solution is computed using a shifted fraction $\tilde{X}_\beta$, which has an equilibrium value of zero at room temperature. Due to the careful formulation of \eqref{eq:analytical_solution_beta}, it is possible to accurately work on a small number below machine precision $\varepsilon(1) \approx \num{1e-16}$ without losing precision by adding to a numeric value on the order of 1.
The analytical solution \eqref{eq:analytical_solution_beta} is evaluated in subsequent time steps until $\tilde{X}_\beta^{n+1}\Delta t > \num{1e-15}$ which is an approximation for the decrement in $\alpha_s$-fraction according to \eqref{eq:alpha_s_to_beta}.
This criterion ensures a sufficient change is noticeable in the $\alpha$-fraction and allows to continue the evolution with \eqref{eq:alpha_s_to_beta}.
While using the approximate analytical solution, the fraction $\tilde{X}_{\beta}$ is tracked independently and not computed from the continuity constraint $\eqref{eq:time_integration_beta_postprocess}$.
The applicability of this strategy has been verified for a constant temperature $T=1200\, \si{\kelvin}$ as shown in Figure~\ref{fig:analytical_solution_xb_verification}. The analytical approximation \eqref{eq:analytical_solution_beta} is exact in the case of a constant temperature.

An analytical approximation for the initial diffusion step from $\beta\rightarrow\alpha_s$ has already been discussed in \cite{nitzler2021novel}. In this case, the values obtained from the analytical expression are large enough and significant digits are not absorbed.

\bibliographystyle{abbrv}
\bibliography{ref}
\end{document}